\documentclass[letter]{aa}
\usepackage{graphicx, graphics}
\usepackage{CJK}
\usepackage{bbm}
\usepackage{amsmath, amssymb, bm}
\usepackage[nodayofweek]{datetime}
\usepackage{txfonts}
\usepackage{placeins}
\usepackage{ulem}
\usepackage[colorlinks=true,citecolor=blue]{hyperref}

\usepackage[switch]{lineno}
\usepackage{tikz}

\definecolor{lime}{HTML}{A6CE39}
\DeclareRobustCommand{\orcidicon}{%
    \raisebox{-3pt}{\begin{tikzpicture}
    \filldraw [lime, yshift=-2pt] (0, 0) circle [radius=0.16]
    node[white] {\raisebox{1pt}{\hspace{0.5pt}\fontfamily{qag}\selectfont\tiny i\scalebox{0.8}{D}}};
    \end{tikzpicture}}
    \hspace{-2.5mm}
    \vspace{-0.25pt}
}

\global\def\tablenotemark#1{{\color{blue}{\normalfont\textsuperscript{\scriptsize #1}}}} 

\newdateformat{monthyearday}{%
  \THEYEAR\ \monthname[\THEMONTH] \twodigit{\THEDAY}}
\newdateformat{daymonthyear}{%
  \twodigit{\THEDAY}\ \monthname[\THEMONTH] \THEYEAR}

\newcommand{\orcidauthor}[2]{#2\href{http://orcid.org/#1}{\orcidicon}}

\titlerunning{Dynamical detection of a companion-driving spiral}
\authorrunning{Xie et al.}

\begin{document}

\begin{CJK*}{UTF8}{gbsn}
\title{Dynamical detection of a companion driving a spiral arm in a protoplanetary disk}

\author{
\orcidauthor{0000-0002-6318-0104}{Chen Xie (谢晨)}\inst{\ref{inst-lam}}
\and
\orcidauthor{0000-0003-1698-9696}{Bin B. Ren (任彬)}\thanks{Marie Sk\l odowska-Curie Fellow}\inst{\ref{inst-oca}, \ref{inst-uga}}
\and
\orcidauthor{0000-0001-9290-7846}{Ruobing Dong  (董若冰)}\inst{\ref{inst-uvic}}
\and
\orcidauthor{0000-0002-9173-0740}{\'Elodie Choquet}\inst{\ref{inst-lam}}
\and
\orcidauthor{0000-0002-5902-7828}{Arthur Vigan}\inst{\ref{inst-lam}}
\and
\orcidauthor{0000-0001-9423-6062}{Jean-Fran\c{c}ois Gonzalez}\inst{\ref{inst-lyon}}
\and
\orcidauthor{0000-0002-4309-6343}{Kevin Wagner}\inst{\ref{inst-UA}}
\and
\orcidauthor{0000-0002-2853-3808}{Taotao Fang  (方陶陶)}\inst{\ref{inst-xmu}}
\and
\orcidauthor{0000-0002-5980-4287}{Maria Giulia Ubeira-Gabellini}\inst{\ref{inst-italy}}
}

\institute{
Aix Marseille Univ, CNRS, CNES, LAM, Marseille, France; \email{\url{chen.xie@lam.fr}}\label{inst-lam}
\and
Universit\'{e} C\^{o}te d'Azur, Observatoire de la C\^{o}te d'Azur, CNRS, Laboratoire Lagrange, Bd de l'Observatoire, CS 34229, 06304 Nice cedex 4, France \label{inst-oca}
\and
Universit\'{e} Grenoble Alpes, CNRS, Institut de Plan\'{e}tologie et d'Astrophysique (IPAG), F-38000 Grenoble, France \label{inst-uga}
\and
Department of Physics \& Astronomy, University of Victoria, Victoria, BC, V8P 5C2, Canada \label{inst-uvic}
\and
Univ Lyon, Univ Claude Bernard Lyon 1, ENS de Lyon, CNRS, Centre de Recherche Astrophysique de Lyon UMR5574, F-69230, Saint-Genis-Laval, France \label{inst-lyon}
\and
Steward Observatory, University of Arizona, USA \label{inst-UA}
\and
Department of Astronomy, Xiamen University, 1 Zengcuoan West Road, Xiamen, Fujian 361005, China \label{inst-xmu}
\and
Dipartimento di Fisica, Universit\`a degli Studi di Milano, Via Celoria 16, 20133 Milano MI, Italy \label{inst-italy}
}

\date{Received \daymonthyear\today; accepted --}

\abstract
{Radio and near-infrared observations have observed dozens of protoplanetary disks that host spiral arm features. 
Numerical simulations have shown that companions may excite spiral density waves in protoplanetary disks via companion--disk interaction. However, the lack of direct observational evidence  for spiral-driving companions poses challenges to current theories of companion--disk interaction. Here we report multi-epoch observations of the binary system \object{HD\,100453} with the Spectro-Polarimetric High-contrast Exoplanet REsearch (SPHERE) facility at the Very Large Telescope. By recovering the spiral features via robustly removing starlight contamination, we measure spiral motion across 4 yr to perform dynamical motion analyses. The spiral pattern motion is consistent with the orbital motion of the eccentric companion. With this first observational evidence of a companion driving a spiral arm among protoplanetary disks, we directly and dynamically confirm the long-standing theory on the origin of spiral features in protoplanetary disks. With the pattern motion of companion-driven spirals being independent of companion mass,  here we establish a feasible way of searching for hidden spiral-arm-driving planets that are beyond the detection of existing ground-based high-contrast imagers.}

\maketitle

\section{Introduction} 
\label{sec:intro}
The detection of spiral structures in protoplanetary disks has called for the understanding of spiral formation mechanisms \citep{Dong2018}. Theoretical and hydrodynamical simulation studies have suggested that companion--disk interaction and disk gravitational instability (GI), together with other mechanisms such as vortex and shadowing \citep[e.g.,][]{Marr22, Montesinos2016}, are the most compelling approaches to contest for explaining the origin of spirals. To test spiral formation mechanisms, hydrodynamical simulations \citep[e.g.,][]{Dong2016, Meru2017, Hall2018} and multi-epoch imaging studies \citep[e.g.,][]{Ren2020_MWC758, Boccaletti2021, Xie_Chengyan2021, Safonov2022} have been employed to associate spiral configuration or motion with the  formation mechanism.

Although motion studies can distinguish theoretical mechanisms between companion-driven and GI-induction, there has been no observationally direct dynamical evidence of the co-motion of a companion and the spiral that it drives. This calls for the verification of the basic assumption in associating the companion--disk interaction theory with spiral motion: the co-motion of companion and spiral \citep[e.g.,][]{Ren2018_MWC758}. It is thus of importance to push beyond identifying companions in spiral systems \citep[e.g.,][]{Currie22_ABAur} by directly validating the motion measurement approach. Therefore, we need to apply motion studies to spiral systems with known companion(s).

In addition to validating the formation and motion mechanism of companion-driven spirals, an application to the known companion--spiral system(s) can also formally establish the existence of these currently hidden spiral-driving planets, especially since   such planets are the most compelling targets for confirmation with direct imaging using state-of-the-art telescopes (e.g., VLT/ERIS, \textit{JWST}) in this current era where targeted imaging approaches (e.g., \citealp{Bohn2020_YSES, Currie22, Franson2023_AFLepb, DeRosa2023_AFLepb, Mesa2023_AFLepb}) instead of blind search are necessary to efficiently populate the family of directly imaged exoplanets. To establish the motion pattern in systems with known spirals and companions, multi-epoch imaging using identical instrument and observation modes can be ideal for minimizing instrument differences and data reduction bias.

\object{HD\,100453} is a binary system \citep{Chen2006} at a distance of $103.8 \pm 0.2$~pc \citep{gaiadr3} with an age of $6.5 \pm 0.5$ Myr \citep{Vioque2018}. The primary star \object{HD\,100453\,A} (hereafter HD\,100453) is a young Herbig A9Ve star with a mass of $1.70\pm0.09$ M$_{\odot}$ \citep{Dominik2003}. The protoplanetary disk around the primary star was directly imaged at near-infrared (NIR) and  submillimeter wavelengths, showing 
a cavity, a ring, and two spiral arms from inside out
\citep{Benisty2017, Wagner2018, vanderPlas2019, Rosotti2020}. NIR interferometric observations revealed the presence of the inner disk \citep{Menu2015, Kluska2020, Bohn2022} inside the cavity that is misaligned with and shadows the outer spiral disk \citep{Bohn2022,Benisty2017}.

The secondary star \object{HD\,100453\,B} (hereafter the companion) is an early M star with a mass of $0.2 \pm 0.04$ M$_{\odot}$ \citep{Collins2009}, located at a projected distance of around 1.05\arcsec~(109 au). 
The HD\,100453 system offers a particularly decisive test of co-motion between spirals and companions that might be driving them. Numerical simulations suggest that such a companion can truncate the disk and excite two spiral arms in the remaining disk as observed in the NIR \citep{Dong2016}. The companion-driven origin of the spiral arms was supported by one of the observed spiral arms in the $^{12}$CO line that connects to the companion position by assuming a coplanar orbit \citep{Rosotti2020}. However, the potential mutual inclination between the companion orbit and the spiral disk \citep{vanderPlas2019, Gonzalez2020} raised the possibility of a projection effect. With HD 100453 being the exemplary configuration of a companion--spiral system, we study the motions of the spiral(s) and the companion here.

\section{Observations and data reduction} 
\label{sec:methods}
The HD\,100453 system was observed with the  Spectro-Polarimetric High-contrast Exoplanet REsearch \citep[SPHERE;][]{Beuzit2019} instrument at the Very Large Telescope (VLT) using the InfraRed Dual-band Imager and Spectrograph \citep[IRDIS;][]{IRDIS_Dohlen2008SPIE}. For spiral motion analysis, we retrieved three total intensity observations in K1-band ($\lambda= 2.11~\mu$m and $\Delta \lambda = 0.10~\mu$m) in April 2015 and April 2019, with a time span of 4.0 years. The observation and the pre-processing of the data are summarized in Appendix~\ref{appendix:observation_log}. Throughout the paper the observation of the disk on 08 April 2019 is only used for uncertainty estimation because its integration time is shorter than the observation on 07 April  2019  (see Table~\ref{table:obs_log}).

We processed the calibrated data by applying reference-star differential imaging (RDI) using the reference images assembled from \citet{xie22}. We constructed a coronagraphic model of stellar signals and speckles via data imputation using sequential nonnegative matrix factorization \citep[DI-sNMF;][]{Ren2020_DIsNMF} (see Appendix~\ref{appendix:RDI-DIsNMF} for a description of the detailed procedures).  Combining the two techniques, RDI-DIsNMF is optimized for the direct imaging of circumstellar disks in total intensity by minimizing self-subtraction and overfitting that has plagued previous methods. We first generated the speckle features (i.e., NMF components of the stellar coronagraphic model) based on the disk-free reference library from \citet{xie22} using NMF from \citet{Ren2018_NMF}, then used these features to remove the speckles in HD 100453 observations. 

To avoid the  overfitting problem for RDI \citep[e.g.,][]{Soummer2012, pueyo16} that can change the morphology of spirals and bias spiral motion measurement, we masked out regions that host disk signals in HD~100453 data, and modeled the rest of the region using the NMF components, and then imputed the signals in disk hosting regions \citep{Ren2020_DIsNMF}. With RDI-DIsNMF using well-chosen reference images, 
we were able to  accurately recover the disk morphology with theoretically minimum post-processing artifacts, which is an essential requirement for accurate measurement of the pattern motion of spiral features. 

\section{Dynamical analysis} 
\label{sec:Analysis}

\subsection{Pattern motion of the spiral arm S1}
\label{subsec:motion_measurement_S1}

\begin{figure}[htb!]
\centering
        \includegraphics[width=0.49\textwidth]{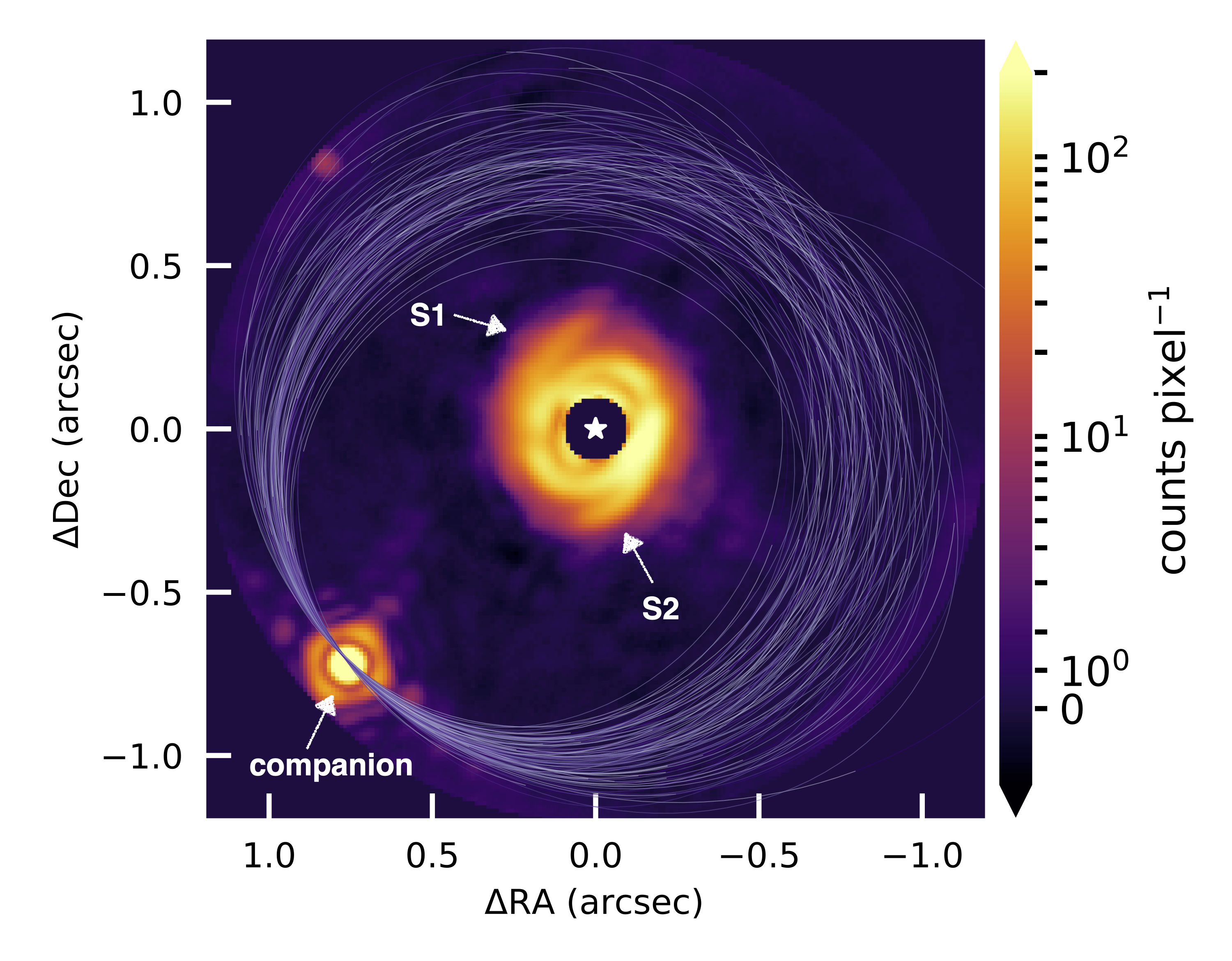}
    \caption{SPHERE/IRDIS detection of the HD100453 system in K1, showing the companion and two spiral arms around a ring-like structure. The gray lines represent the companion orbits that can dynamically drive the spiral arm. The white star shows the position of the primary star.  }
    \label{fig1}
\end{figure}

Using RDI-DIsNMF, we recovered the disk around HD\,100453 at 2.11~$\mu$m in total intensity (Fig.~\ref{fig1}). The disk morphology is consistent with that in the polarimetric image \citep{Benisty2017}. In particular,  we confidently recovered two spiral arms, S1 and S2. S1 is the primary arm that has CO gas connected to the projected position of the companion \citep{Rosotti2020}. To measure the positions of the spiral arms, we first needed to correct the viewing geometry. We deprojected each disk image to face-on views (i.e., the disk plane) adopting an inclination of $33\fdg81$ and a position angle (PA) of $144\fdg35$ \citep{Bohn2022}. To correct for disk flaring, we assumed that  the disk scale height ($h$) follows $h = 0.22 \times r^{1.04}$, where $r$ is radial separation in au \citep{Benisty2017}. Each disk image was then $r^{2}$-scaled to enhance the disk features at large radii. Finally, we transformed them into polar coordinates for the measurement of spiral arm locations. We determined the local maxima of the arm at each azimuthal angle in 1$^{\circ}$ steps by performing Gaussian profile fitting (see Appendix~\ref{appendix:arm_location_fit} for a detailed description). 

\begin{figure}[htb!]
\centering
        \includegraphics[width=0.49\textwidth]{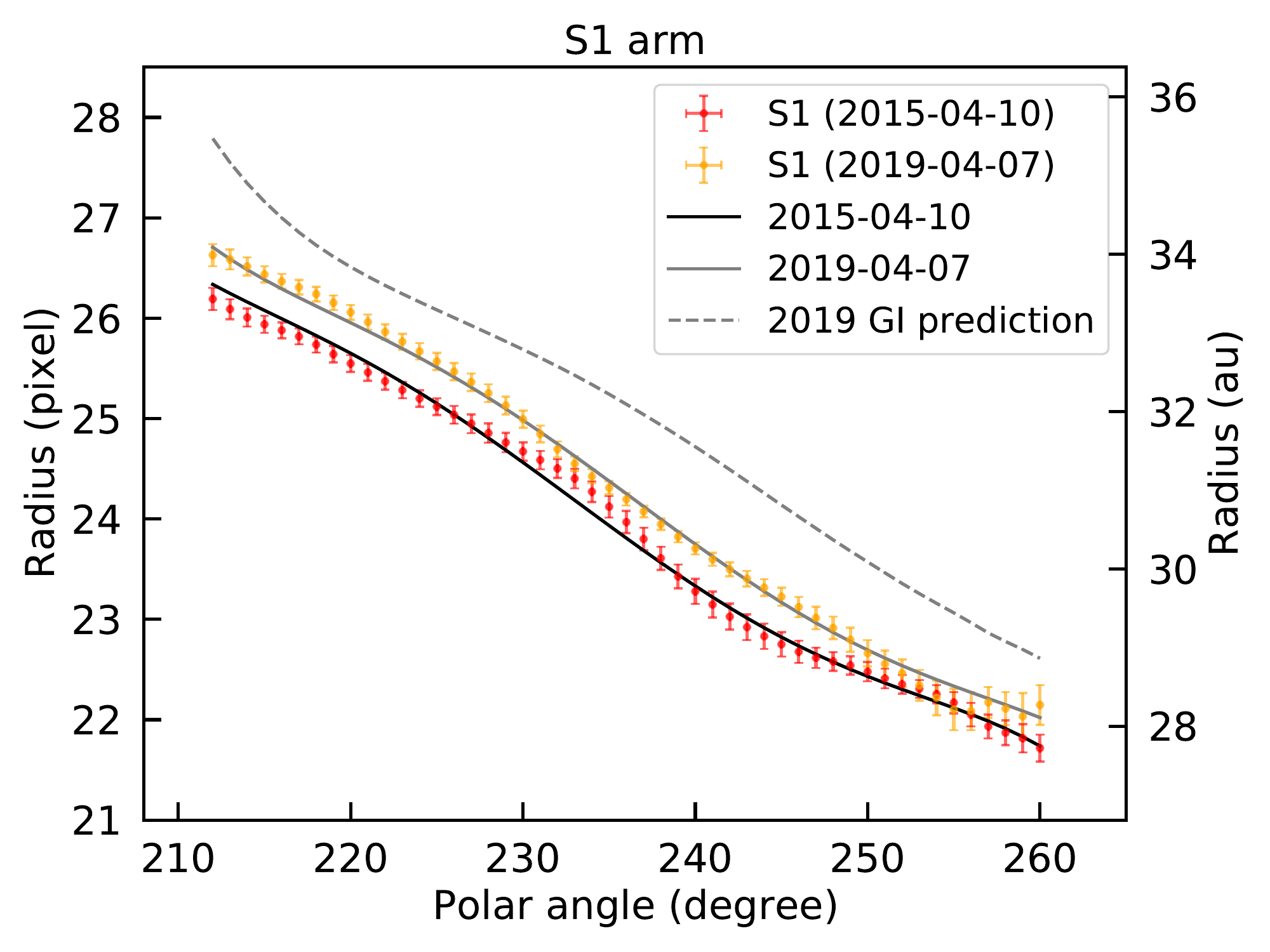}
    \caption{Peak locations of spiral arm S1 in polar coordinates after  correction for  viewing geometry. The solid curves represent the  best-fit model spiral for the peak locations (dot points) between 2015 and 2019, assuming the companion-driven scenario. The derived angular velocities of spiral pattern motion are $0\fdg88 \pm 0\fdg07$~yr$^{-1}$. 
    The uncertainty estimation is presented in Appendix~\ref{appendix:uncertainty_estimation}. The 2019 observations significantly deviate from the gravitational instability prediction (dashed curve), adopting a central star of 1.7~M$_{\odot}$. }
    \label{check_fit_motion_NE}
\end{figure}

The local maxima of spiral arm S1 are presented in Fig.~\ref{check_fit_motion_NE}, which describes the morphology of S1. An offset over a large range of 210$^{\circ}$ -- 260$^{\circ}$ is visible between the two epochs of 2015 and 2019, and follows the rotational direction of the disk. 
This offset also appears when simply comparing disk images between 2015 and 2019 (see  Fig.~\ref{appendix_fig1}).  At locations \textless210$^{\circ}$ we are limited by the signal-to-noise ratio (S/N)  of the disk where the spiral arm is barely detected. At locations \textgreater260$^{\circ}$, the spiral arm starts to merge with the ring-like structure, which increases the uncertainty of the local maxima of the spiral arm. Because possible systematics such as the image misalignment caused by the instrument have been properly corrected during the image alignment (Appendix~\ref{appendix:observation_log}), we conclude that the offset is caused by the pattern motion of the S1 arm in 2015--2019.

Following \cite{Ren2020_MWC758}, we fitted five-degree polynomials to the spiral arm S1 in two epochs and simultaneously obtain their morphological parameters and the pattern motion of the spiral arms. The fitting result is shown in Fig.~\ref{check_fit_motion_NE}. In principle, different parts of the spiral have different pattern speeds if  driven by a companion on an eccentric orbit. However, our data do not permit a radius-dependent assessment, and a single-value pattern speed is measured as a compromise. The speed of the pattern motion for spiral arm S1 is $0\fdg88$ $\pm 0\fdg07$~yr$^{-1}$ in the counterclockwise direction, assuming the companion-driven scenario that the pattern speed is a constant for S1. The uncertainty estimation is described in Appendix~\ref{appendix:uncertainty_estimation}. The deprojection will affect the spiral location determination, and subsequently the spiral motion measurement. However, the disk flaring of HD\,100453 only has limited impact on the velocity of the spiral motion (see Appendix~\ref{appendix:disk_scale_height}). Throughout the paper we present the velocity of the spiral motion based on the best-fit model ($h = 0.22 \times r^{1.04}$) from \cite{Benisty2017} to correct for the disk flaring in the deprojection.

We examined the possibility of the gravitational instability (GI) scenario that each part of the arms rotates at its local Keplerian motion. Based on the fitted morphological parameters of the S1 arm in 2015, we predicted the location of the S1 arm in 2019,  and  adopted the stellar mass of 1.7~M$_{\odot}$ for the central star. The predicted locations of GI-induced spiral arms   deviate from the observed arm locations in 2019 (Fig.~\ref{check_fit_motion_NE}). The local Keplerian motions at 25 to 35 au are   $3\fdg75$~yr$^{-1}$ to $2\fdg27$~yr$^{-1}$, or $15\fdg01$ to $9\fdg06$ in 4 yr. The local Keplerian motions are too large to explain the observation ($\sim$3.5$^\circ$ in 4 years). We conclude that the spiral arm S1 is not triggered by gravitational instability. Together with \object{MWC\,758} \citep{Ren2020_MWC758} and \object{SAO\,206462} \citep{Xie_Chengyan2021}, HD\,100453 is the third spiral disk that disfavors the GI origin for their spirals from pattern motion measurement. 

\begin{figure*}[htb!]
    \centering
        \includegraphics[width=0.95\textwidth]{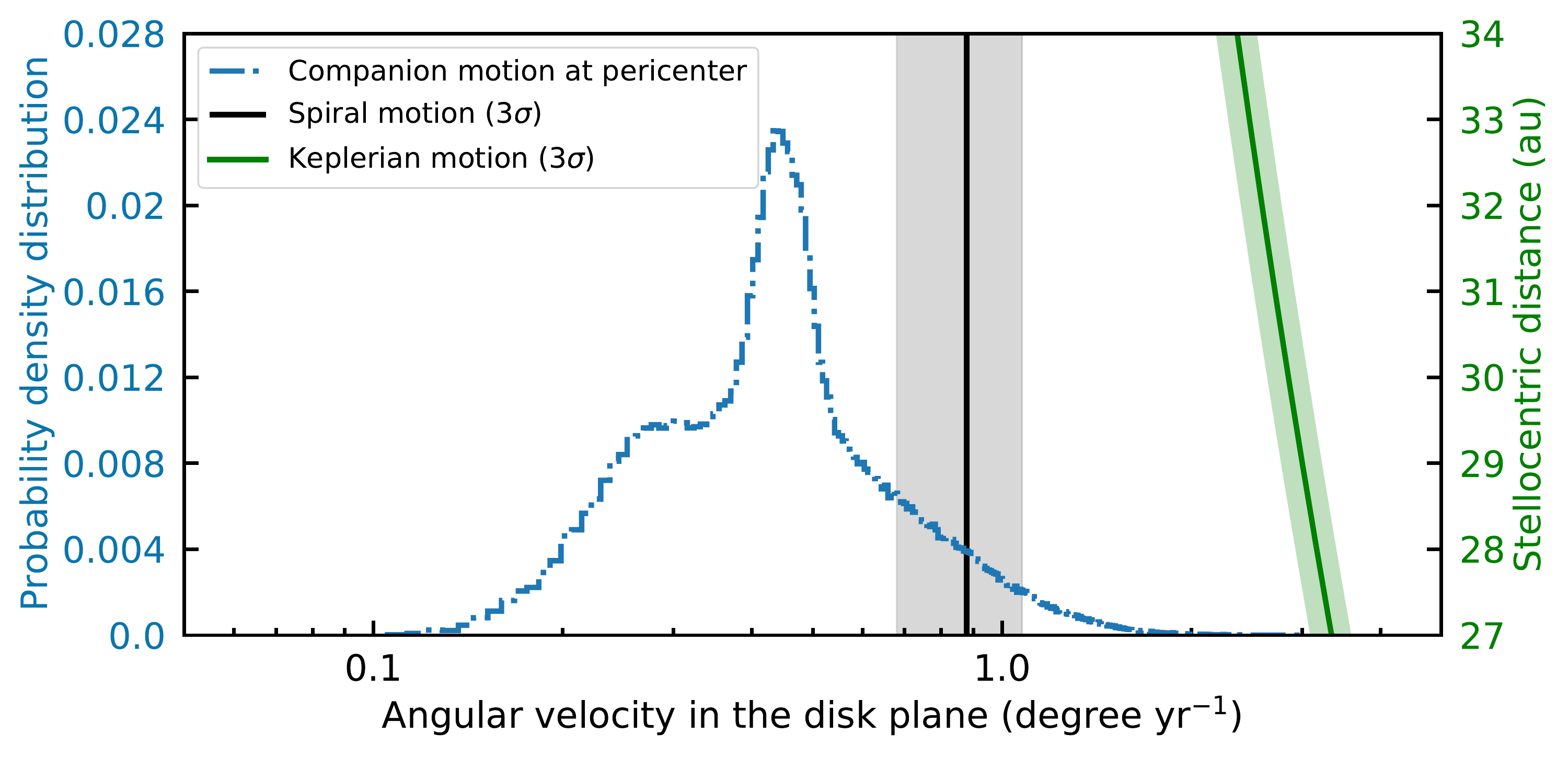}
    \caption{Angular velocity of the S1 spiral pattern motion in HD 100453 (black line). It could follow local Keplerian motion (green curve) or be smaller than the maximum value of the companion orbital motion (blue curve), under gravitational instability (GI) or companion--disk interaction, respectively. The measured spiral motion is consistent with the companion motion in the system and is inconsistent with the local Keplerian motion, revealing the companion-driven origin of the spiral arm S1, while excluding the GI origin. The motions here adopt a central stellar mass of $1.70\pm0.09$ M$_{\odot}$.  The colored shaded regions around each curve represent the corresponding uncertainties.}
    \label{fig2}
\end{figure*}

\subsection{Motion of the stellar companion}

The companion \object{HD\,100453\,B}  has over 15 years of astrometric data. We adopt astrometric data from  \cite{Collins2009}, \cite{Wagner2018}, and \cite{Gonzalez2020} (see Table~\ref{table:astrometry}). A linear fit to the PAs of the companion shows that the companion has an angular velocity of $0\fdg384 \pm 0\fdg019$~yr$^{-1}$ in the sky plane between 2003 and 2019 (Appendix~\ref{appendix:linear_fit_orbital_motion}). Based on the PAs only in 2015 and 2019, we also obtain an angular velocity of $0\fdg40 \pm 0\fdg07$~yr$^{-1}$ for the companion, which is consistent with our linear fit. Hence, we adopt the fitting result as the angular velocity of the companion because it contains more independent measurements. To obtain the companion motion in the disk plane, we deprojected the angular velocity of the companion from the sky plane to the disk plane. In the deprojection, we adopt the disk inclination, the disk position angle, and the companion position angle to be $33\fdg81$, $144\fdg35$, and $133\fdg2$, respectively. We obtain the angular velocity of the companion in the disk plane to be $0\fdg455 \pm 0\fdg023$~yr$^{-1}$ between 2015 and 2019.

We also estimate the probability density distribution of the companion angular velocity in the disk plane (Appendix~\ref{appendix:pdf_companion_orbit}). We calculate the companion angular velocity based on the companion orbital parameters adopted from the orbit fitting results in \cite{Gonzalez2020}. The derived probability density distribution has a Gaussian profile;  the companion has an angular velocity of $0\fdg457^{+0\fdg023}_{-0\fdg023}$~yr$^{-1}$ (1$\sigma$ credible interval) in the disk plane in 2019. The angular velocity derived from the orbital parameters is consistent with the direct linear fit to the companion PAs. This consistent angular velocity and its Gaussian distribution are expected  because the companion motion between 2003 and 2019 is well constrained by the astrometry data with the measuring uncertainty followed potential Gaussian noise. 

In general, the spiral pattern motion should be in the range of the slowest and fastest companion orbital frequency in the scenario of an eccentric perturber \citep[see Eq.~12 in][]{Zhu2022}. From the posterior probability distribution of the orbital parameters obtained by \cite{Gonzalez2020}, we derived the minimum and maximum value of the companion orbital motion. 
Although the companion motion at the location in 2019 is slower than the spiral motion in 2019, the maximum value of the companion orbital motion is still larger than the measured spiral motion  (see Fig.~\ref{fig2}). It suggests that the physical interaction (i.e., tidal interaction) between the companion and the disk can exist, as proposed by the numerical simulation in \citet{Dong2016}. 

Based on the consistency in motion measurements, we conclude that the known companion HD\,100453\,B drives the spiral arm S1. This is the first detection of a companion driving a spiral arm among protoplanetary disks. In light of our result, the previously observed CO gas extending from the S1 arm in \citet{Rosotti2020} is also dynamically connected to the companion, rather than moving independently, and the static connection arises from projection effects due to the relative inclination between the disk and the companion orbit.

\section{Discussion}

\subsection{Pattern motion of the spiral arm S2}

HD\,100453 was classified as a Group I disk that has an outer disk flaring \citep{Meeus2001}. The bottom of the S2 spiral arm shown in the NIR polarimetric image suggests the disk thickness is nonnegligible and the S2 arm locates on the near side of the disk \citep{Benisty2017}. NIR observations probe the scattered light from the disk surface. For the disk with flaring, the angle between the disk surface on the nearside of the disk (i.e., S2) and the sky plane is larger than the disk inclination of $33\fdg8$. We define the disk inclination to be the angle between the flat disk midplane and the sky plane. Large viewing angles (i.e., \textgreater30\fdg0) prevent us from restoring the correct face-on morphology via the deprojection \citep{Dong2016_viewing_angles}. Unlike S2, S1 locates on the far side of the disk where the inclination of the disk surface is smaller than the disk inclination. Thus, it is more accurate to restore the S1 arm back to face-on morphology than the S2 arm. Therefore, we did not provide a motion measurement for the S2 arm. 

\subsection{Other spiral-triggering mechanisms}
\label{subsec:other_mechanisms}
Spiral features can also be induced by a flyby \citep{Menard2020, Dong2022NatAs}. However, a recent search for potential stellar flybys with Gaia DR3 did not identify recent close-in on-sky flyby candidates around HD\,100453 \citep{Shuai2022}. Thus, this scenario is unlikely to cause the spiral feature in this system. 

The disk of HD\,100453 seen in the NIR contains two shadows (Fig.~\ref{fig1}) created by the inner disk \citep{Bohn2022}. \cite{Benisty2017} explored the possibility of the shadow-triggered spirals via the pressure decrease \citep{Montesinos2016}. However, spiral arms triggered by shadows should follow the local Keplerian motion as GI-induced spirals. Therefore, we excluded the shadows being the origin of the spiral features in HD\,100453.

\subsection{Feasibility of locating spiral-driving planets}

In protoplanetary and transitional disks, the occurrence rate and orbital distribution of the embedded planet population remain to be established due to a limited number of confirmed proto-planets.
The current high-contrast imagers with low spectrum resolution (i.e., $R$\textless100) at NIR wavelengths cannot easily discriminate between the scattered light from the dusty disk and that of the embedded planet \citep{Rameau2017, Currie2019, Rich2019}.  H$\alpha$ surveys for the signal of planetary accretions mostly resulted in nondetections of planets \citep{Cugno2019, Zurlo2020, Xie2020}, possibly caused by high extinction or periodic accretions if the forming planet is present. In summary, current instruments with conventional techniques are inefficient in the search for planets embedded in disks. 

Our dynamical motion analysis of the HD\,100453 system validated the approach of mapping spiral arm motion to locate hidden giant planets in protoplanetary disks, first proposed by \cite{Ren2020_MWC758}. Although we only investigated a nonplanetary companion in this specific study, the pattern speed of companion-driven spirals depends on the location of the companion instead of its mass. Furthermore, the sensitivity of our motion measurement directly depends on the time span (see Eq.~\eqref{eq:uncertainty}). 
The uncertainty of the motion measurement decreases with the increase in  the time span ($t$) of two epochs. Typical 1$\sigma$ uncertainties for the motion measurements based on two epochs (if $t$=5 yr) of SPHERE observations in total intensity and polarized light are about $0\fdg05$~yr$^{-1}$ (Appendix~\ref{appendix:uncertainty_estimation}) and $0\fdg03$~yr$^{-1}$ \citep{Ren2020_MWC758}, respectively. Therefore, spiral motions driven by a planet can be detected and distinguished from local Keplerian motions (GI scenario) within a feasible time of a few years. 

From the posterior probability distribution of orbital parameters obtained by \cite{Gonzalez2020}, we derived the companion orbits that can dynamically drive the spiral arm (see Fig.~\ref{fig1}). The corresponding probability distribution of orbital parameters is shown in Fig.~\ref{appendix_cornor_plot}.  Our dynamical analysis opens a new and feasible window to probe the orbit distribution of planets in the spiral disks that currently are difficult to study via conventional techniques of direct imaging and spectroscopy. The measured spiral motion can determine the range of the planet eccentricity \citep[e.g., Eq.12 in][]{Zhu2022}. In combination with the planet mass estimated from the morphology of the spiral arms \citep{Fung2015} or simply using mass upper limits from direct imaging, we can infer the formation and migration of planets at the early stage.

\section{Conclusion}

We present multi-epoch observations of the HD\,100453 system and perform a dynamical analysis for the spiral motion in 4 yr. The measured pattern motion of the spiral arm S1 disfavors the GI origin. More importantly, the orbital motion of companion HD\,100453\,B can explain the spiral pattern motion in the scenario of the eccentric perturber. It is the first dynamical detection of a companion driving a spiral arm among protoplanetary disks.

Companion--disk interaction is a long-standing theory that could naturally explain the origin of spiral features in disks. 
For the first time, our dynamical analyses directly confirm that the companion--disk interaction can indeed induce spiral arms in disks, supporting that it could also be the formation mechanism for other spiral systems without detected companions. Our dynamical detection also validates our method to be a feasible way of searching for and locating hidden spiral-arm-driving planets that are best targets for dedicated direct imaging explorations \citep[e.g.,][Fig.~7 therein]{Bae22} with upcoming state-of-the-art high-contrast imagers. 

\begin{acknowledgements}
We thank Dr.~Faustine~Cantalloube for the beneficial discussion about the low wind effect. The disk images of HD\,100453 are based on observations collected at the European Organisation for Astronomical Research in the Southern Hemisphere under ESO programmes 095.C-0389(A) and 0103.C-0847(A). We thank all the principal investigators and their collaborators who prepared and performed the observations with SPHERE. Without their efforts, we would not be able to build the master reference library to enable our RDI technique. B.B.R.~acknowledges funding from the European Research Council (ERC) under the European Union's Horizon 2020 research and innovation programme (grant PROTOPLANETS No.~101002188), and from the European Union's Horizon 2020 research and innovation programme under the Marie Sk\l odowska-Curie grant agreement No.~101103114. R.D.~acknowledges financial support provided by the Natural Sciences and Engineering Research Council of Canada through a Discovery Grant, as well as the Alfred P.~Sloan Foundation through a Sloan Research Fellowship.  E.C. acknowledges funding from the European Research Council (ERC) under the European Union's Horizon Europe research and innovation programme (ESCAPE, grant agreement No 101044152). A.V.~acknowledges funding from the European Research Council (ERC) under the European Union's Horizon 2020 research and innovation programme (grant agreement No.~757561). J.-F.G.~acknowledges funding from the European Union's Horizon 2020 research and innovation programme under the Marie Sk\l{}odowska-Curie grant agreement No 823823 (DUSTBUSTERS) and from the ANR (Agence Nationale de la Recherche) of France under contract number ANR-16- CE31-0013 (Planet-Forming-Disks), and thanks the LABEX Lyon Institute of Origins (ANR-10-LABX-0066) for its financial support within the Plan France 2030 of the French government operated by the ANR. T.F.~acknowledges supports from the National Key R\&D Program of China No.~2017YFA0402600, from NSFC grants No.~11890692, 12133008, and 12221003, and from the science research grants from the China Manned Space Project with No.~CMS-CSST-2021-A04. 

\end{acknowledgements}

\bibliographystyle{aa}
\bibliography{ms}

\begin{thebibliography}{67}
\expandafter\ifx\csname natexlab\endcsname\relax\def\natexlab#1{#1}\fi

\bibitem[{{Bae} {et~al.}(2022){Bae}, {Isella}, {Zhu}, {Martin}, {Okuzumi}, \&
  {Suriano}}]{Bae22}
{Bae}, J., {Isella}, A., {Zhu}, Z., {et~al.} 2022, arXiv e-prints,
  arXiv:2210.13314

\bibitem[{{Benisty} {et~al.}(2017){Benisty}, {Stolker}, {Pohl}, {de Boer},
  {Lesur}, {Dominik}, {Dullemond}, {Langlois}, {Min}, {Wagner}, {Henning},
  {Juhasz}, {Pinilla}, {Facchini}, {Apai}, {van Boekel}, {Garufi}, {Ginski},
  {M{\'e}nard}, {Pinte}, {Quanz}, {Zurlo}, {Boccaletti}, {Bonnefoy}, {Beuzit},
  {Chauvin}, {Cudel}, {Desidera}, {Feldt}, {Fontanive}, {Gratton}, {Kasper},
  {Lagrange}, {LeCoroller}, {Mouillet}, {Mesa}, {Sissa}, {Vigan}, {Antichi},
  {Buey}, {Fusco}, {Gisler}, {Llored}, {Magnard}, {Moeller-Nilsson}, {Pragt},
  {Roelfsema}, {Sauvage}, \& {Wildi}}]{Benisty2017}
{Benisty}, M., {Stolker}, T., {Pohl}, A., {et~al.} 2017, \aap, 597, A42

\bibitem[{{Beuzit} {et~al.}(2019){Beuzit}, {Vigan}, {Mouillet}, {Dohlen},
  {Gratton}, {Boccaletti}, {Sauvage}, {Schmid}, {Langlois}, {Petit},
  {Baruffolo}, {Feldt}, {Milli}, {Wahhaj}, {Abe}, {Anselmi}, {Antichi},
  {Barette}, {Baudrand}, {Baudoz}, {Bazzon}, {Bernardi}, {Blanchard}, {Brast},
  {Bruno}, {Buey}, {Carbillet}, {Carle}, {Cascone}, {Chapron}, {Charton},
  {Chauvin}, {Claudi}, {Costille}, {De Caprio}, {de Boer}, {Delboulb{\'e}},
  {Desidera}, {Dominik}, {Downing}, {Dupuis}, {Fabron}, {Fantinel}, {Farisato},
  {Feautrier}, {Fedrigo}, {Fusco}, {Gigan}, {Ginski}, {Girard}, {Giro},
  {Gisler}, {Gluck}, {Gry}, {Henning}, {Hubin}, {Hugot}, {Incorvaia}, {Jaquet},
  {Kasper}, {Lagadec}, {Lagrange}, {Le Coroller}, {Le Mignant}, {Le Ruyet},
  {Lessio}, {Lizon}, {Llored}, {Lundin}, {Madec}, {Magnard}, {Marteaud},
  {Martinez}, {Maurel}, {M{\'e}nard}, {Mesa}, {M{\"o}ller-Nilsson}, {Moulin},
  {Moutou}, {Orign{\'e}}, {Parisot}, {Pavlov}, {Perret}, {Pragt}, {Puget},
  {Rabou}, {Ramos}, {Reess}, {Rigal}, {Rochat}, {Roelfsema}, {Rousset}, {Roux},
  {Saisse}, {Salasnich}, {Santambrogio}, {Scuderi}, {Segransan}, {Sevin},
  {Siebenmorgen}, {Soenke}, {Stadler}, {Suarez}, {Tiph{\`e}ne}, {Turatto},
  {Udry}, {Vakili}, {Waters}, {Weber}, {Wildi}, {Zins}, \&
  {Zurlo}}]{Beuzit2019}
{Beuzit}, J.~L., {Vigan}, A., {Mouillet}, D., {et~al.} 2019, \aap, 631, A155

\bibitem[{{Blunt} {et~al.}(2017){Blunt}, {Nielsen}, {De Rosa}, {Konopacky},
  {Ryan}, {Wang}, {Pueyo}, {Rameau}, {Marois}, {Marchis}, {Macintosh},
  {Graham}, {Duch{\^e}ne}, \& {Schneider}}]{orbitize_Blunt2017}
{Blunt}, S., {Nielsen}, E.~L., {De Rosa}, R.~J., {et~al.} 2017, \aj, 153, 229

\bibitem[{{Blunt} {et~al.}(2020){Blunt}, {Wang}, {Angelo}, {Ngo}, {Cody}, {De
  Rosa}, {Graham}, {Hirsch}, {Nagpal}, {Nielsen}, {Pearce}, {Rice}, \&
  {Tejada}}]{orbitize_Blunt2020}
{Blunt}, S., {Wang}, J.~J., {Angelo}, I., {et~al.} 2020, \aj, 159, 89

\bibitem[{{Boccaletti} {et~al.}(2021){Boccaletti}, {Pantin}, {M{\'e}nard},
  {Galicher}, {Langlois}, {Benisty}, {Gratton}, {Chauvin}, {Ginski},
  {Lagrange}, {Zurlo}, {Biller}, {Bonavita}, {Bonnefoy}, {Brown-Sevilla},
  {Cantalloube}, {Desidera}, {D'Orazi}, {Feldt}, {Hagelberg}, {Lazzoni},
  {Mesa}, {Meyer}, {Perrot}, {Vigan}, {Sauvage}, {Ramos}, {Rousset}, \&
  {Magnard}}]{Boccaletti2021}
{Boccaletti}, A., {Pantin}, E., {M{\'e}nard}, F., {et~al.} 2021, \aap, 652, L8

\bibitem[{{Bohn} {et~al.}(2022){Bohn}, {Benisty}, {Perraut}, {van der Marel},
  {W{\"o}lfer}, {van Dishoeck}, {Facchini}, {Manara}, {Teague}, {Francis},
  {Berger}, {Garcia-Lopez}, {Ginski}, {Henning}, {Kenworthy}, {Kraus},
  {M{\'e}nard}, {M{\'e}rand}, \& {P{\'e}rez}}]{Bohn2022}
{Bohn}, A.~J., {Benisty}, M., {Perraut}, K., {et~al.} 2022, \aap, 658, A183

\bibitem[{{Bohn} {et~al.}(2020){Bohn}, {Kenworthy}, {Ginski}, {Manara},
  {Pecaut}, {de Boer}, {Keller}, {Mamajek}, {Meshkat}, {Reggiani}, {Todorov},
  \& {Snik}}]{Bohn2020_YSES}
{Bohn}, A.~J., {Kenworthy}, M.~A., {Ginski}, C., {et~al.} 2020, \mnras, 492,
  431

\bibitem[{{Cantalloube} {et~al.}(2020){Cantalloube}, {Farley}, {Milli},
  {Bharmal}, {Brandner}, {Correia}, {Dohlen}, {Henning}, {Osborn}, {Por},
  {Su{\'a}rez Valles}, \& {Vigan}}]{Cantalloube2020}
{Cantalloube}, F., {Farley}, O.~J.~D., {Milli}, J., {et~al.} 2020, \aap, 638,
  A98

\bibitem[{{Carbillet} {et~al.}(2011){Carbillet}, {Bendjoya}, {Abe}, {Guerri},
  {Boccaletti}, {Daban}, {Dohlen}, {Ferrari}, {Robbe-Dubois}, {Douet}, \&
  {Vakili}}]{Carbillet2011}
{Carbillet}, M., {Bendjoya}, P., {Abe}, L., {et~al.} 2011, Experimental
  Astronomy, 30, 39

\bibitem[{{Chauvin} {et~al.}(2010){Chauvin}, {Lagrange}, {Bonavita},
  {Zuckerman}, {Dumas}, {Bessell}, {Beuzit}, {Bonnefoy}, {Desidera}, {Farihi},
  {Lowrance}, {Mouillet}, \& {Song}}]{Chauvin2010}
{Chauvin}, G., {Lagrange}, A.~M., {Bonavita}, M., {et~al.} 2010, \aap, 509, A52

\bibitem[{{Chen} {et~al.}(2006){Chen}, {Henning}, {van Boekel}, \&
  {Grady}}]{Chen2006}
{Chen}, X.~P., {Henning}, T., {van Boekel}, R., \& {Grady}, C.~A. 2006, \aap,
  445, 331

\bibitem[{{Collins} {et~al.}(2009){Collins}, {Grady}, {Hamaguchi},
  {Wisniewski}, {Brittain}, {Sitko}, {Carpenter}, {Williams}, {Mathews},
  {Williger}, {van Boekel}, {Carmona}, {Henning}, {van den Ancker}, {Meeus},
  {Chen}, {Petre}, \& {Woodgate}}]{Collins2009}
{Collins}, K.~A., {Grady}, C.~A., {Hamaguchi}, K., {et~al.} 2009, \apj, 697,
  557

\bibitem[{{Cugno} {et~al.}(2019){Cugno}, {Quanz}, {Hunziker}, {Stolker},
  {Schmid}, {Avenhaus}, {Baudoz}, {Bohn}, {Bonnefoy}, {Buenzli}, {Chauvin},
  {Cheetham}, {Desidera}, {Dominik}, {Feautrier}, {Feldt}, {Ginski}, {Girard},
  {Gratton}, {Hagelberg}, {Hugot}, {Janson}, {Lagrange}, {Langlois}, {Magnard},
  {Maire}, {Menard}, {Meyer}, {Milli}, {Mordasini}, {Pinte}, {Pragt},
  {Roelfsema}, {Rigal}, {Szul{\'a}gyi}, {van Boekel}, {van der Plas}, {Vigan},
  {Wahhaj}, \& {Zurlo}}]{Cugno2019}
{Cugno}, G., {Quanz}, S.~P., {Hunziker}, S., {et~al.} 2019, \aap, 622, A156

\bibitem[{{Currie} {et~al.}(2023){Currie}, {Brandt}, {Brandt}, {Lacy},
  {Burrows}, {Guyon}, {Tamura}, {Liu}, {Sagynbayeva}, {Tobin}, {Chilcote},
  {Groff}, {Marois}, {Thompson}, {Murphy}, {Kuzuhara}, {Lawson}, {Lozi}, {Deo},
  {Vievard}, {Skaf}, {Uyama}, {Jovanovic}, {Martinache}, {Kasdin}, {Kudo},
  {McElwain}, {Janson}, {Wisniewski}, {Hodapp}, {Nishikawa}, {He{\l}miniak},
  {Kwon}, \& {Hayashi}}]{Currie22}
{Currie}, T., {Brandt}, G.~M., {Brandt}, T.~D., {et~al.} 2023, Science, 380,
  198

\bibitem[{{Currie} {et~al.}(2022){Currie}, {Lawson}, {Schneider}, {Lyra},
  {Wisniewski}, {Grady}, {Guyon}, {Tamura}, {Kotani}, {Kawahara}, {Brandt},
  {Uyama}, {Muto}, {Dong}, {Kudo}, {Hashimoto}, {Fukagawa}, {Wagner}, {Lozi},
  {Chilcote}, {Tobin}, {Groff}, {Ward-Duong}, {Januszewski}, {Norris},
  {Tuthill}, {van der Marel}, {Sitko}, {Deo}, {Vievard}, {Jovanovic},
  {Martinache}, \& {Skaf}}]{Currie22_ABAur}
{Currie}, T., {Lawson}, K., {Schneider}, G., {et~al.} 2022, Nature Astronomy,
  6, 751

\bibitem[{{Currie} {et~al.}(2019){Currie}, {Marois}, {Cieza}, {Mulders},
  {Lawson}, {Caceres}, {Rodriguez-Ruiz}, {Wisniewski}, {Guyon}, {Brandt},
  {Kasdin}, {Groff}, {Lozi}, {Chilcote}, {Hodapp}, {Jovanovic}, {Martinache},
  {Skaf}, {Lyra}, {Tamura}, {Asensio-Torres}, {Dong}, {Grady}, {Gerard},
  {Fukagawa}, {Hand}, {Hayashi}, {Henning}, {Kudo}, {Kuzuhara}, {Kwon},
  {McElwain}, \& {Uyama}}]{Currie2019}
{Currie}, T., {Marois}, C., {Cieza}, L., {et~al.} 2019, \apjl, 877, L3

\bibitem[{{De Rosa} {et~al.}(2023){De Rosa}, {Nielsen}, {Wahhaj}, {Ruffio},
  {Kalas}, {Peck}, {Hirsch}, \& {Roberson}}]{DeRosa2023_AFLepb}
{De Rosa}, R.~J., {Nielsen}, E.~L., {Wahhaj}, Z., {et~al.} 2023, \aap, 672, A94

\bibitem[{{Dohlen} {et~al.}(2008){Dohlen}, {Langlois}, {Saisse}, {Hill},
  {Origne}, {Jacquet}, {Fabron}, {Blanc}, {Llored}, {Carle}, {Moutou}, {Vigan},
  {Boccaletti}, {Carbillet}, {Mouillet}, \& {Beuzit}}]{IRDIS_Dohlen2008SPIE}
{Dohlen}, K., {Langlois}, M., {Saisse}, M., {et~al.} 2008, in Society of
  Photo-Optical Instrumentation Engineers (SPIE) Conference Series, Vol. 7014,
  Ground-based and Airborne Instrumentation for Astronomy II, ed. I.~S.
  {McLean} \& M.~M. {Casali}, 70143L

\bibitem[{{Dominik} {et~al.}(2003){Dominik}, {Dullemond}, {Waters}, \&
  {Walch}}]{Dominik2003}
{Dominik}, C., {Dullemond}, C.~P., {Waters}, L.~B.~F.~M., \& {Walch}, S. 2003,
  \aap, 398, 607

\bibitem[{{Dong} {et~al.}(2016{\natexlab{a}}){Dong}, {Fung}, \&
  {Chiang}}]{Dong2016_viewing_angles}
{Dong}, R., {Fung}, J., \& {Chiang}, E. 2016{\natexlab{a}}, \apj, 826, 75

\bibitem[{{Dong} {et~al.}(2022){Dong}, {Liu}, {Cuello}, {Pinte},
  {{\'A}brah{\'a}m}, {Vorobyov}, {Hashimoto}, {K{\'o}sp{\'a}l}, {Chiang},
  {Takami}, {Chen}, {Dunham}, {Fukagawa}, {Green}, {Hasegawa}, {Henning},
  {Pavlyuchenkov}, {Pyo}, \& {Tamura}}]{Dong2022NatAs}
{Dong}, R., {Liu}, H.~B., {Cuello}, N., {et~al.} 2022, Nature Astronomy, 6, 331

\bibitem[{{Dong} {et~al.}(2018){Dong}, {Najita}, \& {Brittain}}]{Dong2018}
{Dong}, R., {Najita}, J.~R., \& {Brittain}, S. 2018, \apj, 862, 103

\bibitem[{{Dong} {et~al.}(2016{\natexlab{b}}){Dong}, {Zhu}, {Fung}, {Rafikov},
  {Chiang}, \& {Wagner}}]{Dong2016}
{Dong}, R., {Zhu}, Z., {Fung}, J., {et~al.} 2016{\natexlab{b}}, \apjl, 816, L12

\bibitem[{{Franson} {et~al.}(2023){Franson}, {Bowler}, {Zhou}, {Pearce},
  {Bardalez Gagliuffi}, {Biddle}, {Brandt}, {Crepp}, {Dupuy}, {Faherty},
  {Jensen-Clem}, {Morgan}, {Sanghi}, {Theissen}, {Tran}, \&
  {Wolf}}]{Franson2023_AFLepb}
{Franson}, K., {Bowler}, B.~P., {Zhou}, Y., {et~al.} 2023, arXiv e-prints,
  arXiv:2302.05420

\bibitem[{{Fung} \& {Dong}(2015)}]{Fung2015}
{Fung}, J. \& {Dong}, R. 2015, \apjl, 815, L21

\bibitem[{{Gaia Collaboration} {et~al.}(2022){Gaia Collaboration}, {Vallenari},
  {Brown}, {Prusti}, {de Bruijne}, {Arenou}, {Babusiaux}, {Biermann},
  {Creevey}, {Ducourant}, {Evans}, {Eyer}, {Guerra}, {Hutton}, {Jordi},
  {Klioner}, {Lammers}, {Lindegren}, {Luri}, {Mignard}, {Panem}, {Pourbaix},
  {Randich}, {Sartoretti}, {Soubiran}, {Tanga}, {Walton}, {Bailer-Jones},
  {Bastian}, {Drimmel}, {Jansen}, {Katz}, {Lattanzi}, {van Leeuwen}, {Bakker},
  {Cacciari}, {Casta{\~n}eda}, {De Angeli}, {Fabricius}, {Fouesneau},
  {Fr{\'e}mat}, {Galluccio}, {Guerrier}, {Heiter}, {Masana}, {Messineo},
  {Mowlavi}, {Nicolas}, {Nienartowicz}, {Pailler}, {Panuzzo}, {Riclet}, {Roux},
  {Seabroke}, {Sordo{\o}rcit}, {Th{\'e}venin}, {Gracia-Abril}, {Portell},
  {Teyssier}, {Altmann}, {Andrae}, {Audard}, {Bellas-Velidis}, {Benson},
  {Berthier}, {Blomme}, {Burgess}, {Busonero}, {Busso}, {C{\'a}novas}, {Carry},
  {Cellino}, {Cheek}, {Clementini}, {Damerdji}, {Davidson}, {de Teodoro},
  {Nu{\~n}ez Campos}, {Delchambre}, {Dell'Oro}, {Esquej},
  {Fern{\'a}ndez-Hern{\'a}ndez}, {Fraile}, {Garabato}, {Garc{\'\i}a-Lario},
  {Gosset}, {Haigron}, {Halbwachs}, {Hambly}, {Harrison}, {Hern{\'a}ndez},
  {Hestroffer}, {Hodgkin}, {Holl}, {Jan{\ss}en}, {Jevardat de Fombelle},
  {Jordan}, {Krone-Martins}, {Lanzafame}, {L{\"o}ffler}, {Marchal}, {Marrese},
  {Moitinho}, {Muinonen}, {Osborne}, {Pancino}, {Pauwels}, {Recio-Blanco},
  {Reyl{\'e}}, {Riello}, {Rimoldini}, {Roegiers}, {Rybizki}, {Sarro}, {Siopis},
  {Smith}, {Sozzetti}, {Utrilla}, {van Leeuwen}, {Abbas}, {{\'A}brah{\'a}m},
  {Abreu Aramburu}, {Aerts}, {Aguado}, {Ajaj}, {Aldea-Montero}, {Altavilla},
  {{\'A}lvarez}, {Alves}, {Anders}, {Anderson}, {Anglada Varela}, {Antoja},
  {Baines}, {Baker}, {Balaguer-N{\'u}{\~n}ez}, {Balbinot}, {Balog}, {Barache},
  {Barbato}, {Barros}, {Barstow}, {Bartolom{\'e}}, {Bassilana}, {Bauchet},
  {Becciani}, {Bellazzini}, {Berihuete}, {Bernet}, {Bertone}, {Bianchi},
  {Binnenfeld}, {Blanco-Cuaresma}, {Blazere}, {Boch}, {Bombrun}, {Bossini},
  {Bouquillon}, {Bragaglia}, {Bramante}, {Breedt}, {Bressan}, {Brouillet},
  {Brugaletta}, {Bucciarelli}, {Burlacu}, {Butkevich}, {Buzzi}, {Caffau},
  {Cancelliere}, {Cantat-Gaudin}, {Carballo}, {Carlucci}, {Carnerero},
  {Carrasco}, {Casamiquela}, {Castellani}, {Castro-Ginard}, {Chaoul},
  {Charlot}, {Chemin}, {Chiaramida}, {Chiavassa}, {Chornay}, {Comoretto},
  {Contursi}, {Cooper}, {Cornez}, {Cowell}, {Crifo}, {Cropper}, {Crosta},
  {Crowley}, {Dafonte}, {Dapergolas}, {David}, {David}, {de Laverny}, {De
  Luise}, {De March}, {De Ridder}, {de Souza}, {de Torres}, {del Peloso}, {del
  Pozo}, {Delbo}, {Delgado}, {Delisle}, {Demouchy}, {Dharmawardena}, {Di
  Matteo}, {Diakite}, {Diener}, {Distefano}, {Dolding}, {Edvardsson}, {Enke},
  {Fabre}, {Fabrizio}, {Faigler}, {Fedorets}, {Fernique}, {Fienga}, {Figueras},
  {Fournier}, {Fouron}, {Fragkoudi}, {Gai}, {Garcia-Gutierrez},
  {Garcia-Reinaldos}, {Garc{\'\i}a-Torres}, {Garofalo}, {Gavel}, {Gavras},
  {Gerlach}, {Geyer}, {Giacobbe}, {Gilmore}, {Girona}, {Giuffrida}, {Gomel},
  {Gomez}, {Gonz{\'a}lez-N{\'u}{\~n}ez}, {Gonz{\'a}lez-Santamar{\'\i}a},
  {Gonz{\'a}lez-Vidal}, {Granvik}, {Guillout}, {Guiraud},
  {Guti{\'e}rrez-S{\'a}nchez}, {Guy}, {Hatzidimitriou}, {Hauser}, {Haywood},
  {Helmer}, {Helmi}, {Sarmiento}, {Hidalgo}, {Hilger}, {H{\l}adczuk}, {Hobbs},
  {Holland}, {Huckle}, {Jardine}, {Jasniewicz}, {Jean-Antoine Piccolo},
  {Jim{\'e}nez-Arranz}, {Jorissen}, {Juaristi Campillo}, {Julbe}, {Karbevska},
  {Kervella}, {Khanna}, {Kontizas}, {Kordopatis}, {Korn}, {K{\'o}sp{\'a}l},
  {Kostrzewa-Rutkowska}, {Kruszy{\'n}ska}, {Kun}, {Laizeau}, {Lambert},
  {Lanza}, {Lasne}, {Le Campion}, {Lebreton}, {Lebzelter}, {Leccia}, {Leclerc},
  {Lecoeur-Taibi}, {Liao}, {Licata}, {Lindstr{\o}m}, {Lister}, {Livanou},
  {Lobel}, {Lorca}, {Loup}, {Madrero Pardo}, {Magdaleno Romeo}, {Managau},
  {Mann}, {Manteiga}, {Marchant}, {Marconi}, {Marcos}, {Marcos Santos},
  {Mar{\'\i}n Pina}, {Marinoni}, {Marocco}, {Marshall}, {Polo},
  {Mart{\'\i}n-Fleitas}, {Marton}, {Mary}, {Masip}, {Massari},
  {Mastrobuono-Battisti}, {Mazeh}, {McMillan}, {Messina}, {Michalik}, {Millar},
  {Mints}, {Molina}, {Molinaro}, {Moln{\'a}r}, {Monari}, {Mongui{\'o}},
  {Montegriffo}, {Montero}, {Mor}, {Mora}, {Morbidelli}, {Morel}, {Morris},
  {Muraveva}, {Murphy}, {Musella}, {Nagy}, {Noval}, {Oca{\~n}a}, {Ogden},
  {Ordenovic}, {Osinde}, {Pagani}, {Pagano}, {Palaversa}, {Palicio},
  {Pallas-Quintela}, {Panahi}, {Payne-Wardenaar}, {Pe{\~n}alosa Esteller},
  {Penttil{\"a}}, {Pichon}, {Piersimoni}, {Pineau}, {Plachy}, {Plum}, {Poggio},
  {Pr{\v{s}}a}, {Pulone}, {Racero}, {Ragaini}, {Rainer}, {Raiteri}, {Rambaux},
  {Ramos}, {Ramos-Lerate}, {Re Fiorentin}, {Regibo}, {Richards}, {Rios Diaz},
  {Ripepi}, {Riva}, {Rix}, {Rixon}, {Robichon}, {Robin}, {Robin}, {Roelens},
  {Rogues}, {Rohrbasser}, {Romero-G{\'o}mez}, {Rowell}, {Royer}, {Ruz Mieres},
  {Rybicki}, {Sadowski}, {S{\'a}ez N{\'u}{\~n}ez}, {Sagrist{\`a} Sell{\'e}s},
  {Sahlmann}, {Salguero}, {Samaras}, {Sanchez Gimenez}, {Sanna},
  {Santove{\~n}a}, {Sarasso}, {Schultheis}, {Sciacca}, {Segol}, {Segovia},
  {S{\'e}gransan}, {Semeux}, {Shahaf}, {Siddiqui}, {Siebert}, {Siltala},
  {Silvelo}, {Slezak}, {Slezak}, {Smart}, {Snaith}, {Solano}, {Solitro},
  {Souami}, {Souchay}, {Spagna}, {Spina}, {Spoto}, {Steele},
  {Steidelm{\"u}ller}, {Stephenson}, {S{\"u}veges}, {Surdej}, {Szabados},
  {Szegedi-Elek}, {Taris}, {Taylo}, {Teixeira}, {Tolomei}, {Tonello}, {Torra},
  {Torra}, {Torralba Elipe}, {Trabucchi}, {Tsounis}, {Turon}, {Ulla}, {Unger},
  {Vaillant}, {van Dillen}, {van Reeven}, {Vanel}, {Vecchiato}, {Viala},
  {Vicente}, {Voutsinas}, {Weiler}, {Wevers}, {Wyrzykowski}, {Yoldas}, {Yvard},
  {Zhao}, {Zorec}, {Zucker}, \& {Zwitter}}]{gaiadr3}
{Gaia Collaboration}, {Vallenari}, A., {Brown}, A.~G.~A., {et~al.} 2022, arXiv
  e-prints, arXiv:2208.00211

\bibitem[{{Gonzalez} {et~al.}(2020){Gonzalez}, {van der Plas}, {Pinte},
  {Cuello}, {Nealon}, {M{\'e}nard}, {Revol}, {Rodet}, {Langlois}, \&
  {Maire}}]{Gonzalez2020}
{Gonzalez}, J.-F., {van der Plas}, G., {Pinte}, C., {et~al.} 2020, \mnras, 499,
  3837

\bibitem[{{Guerri} {et~al.}(2011){Guerri}, {Daban}, {Robbe-Dubois}, {Douet},
  {Abe}, {Baudrand}, {Carbillet}, {Boccaletti}, {Bendjoya}, {Gouvret}, \&
  {Vakili}}]{Guerri2011}
{Guerri}, G., {Daban}, J.-B., {Robbe-Dubois}, S., {et~al.} 2011, Experimental
  Astronomy, 30, 59

\bibitem[{{Hall} {et~al.}(2018){Hall}, {Rice}, {Dipierro}, {Forgan}, {Harries},
  \& {Alexander}}]{Hall2018}
{Hall}, C., {Rice}, K., {Dipierro}, G., {et~al.} 2018, \mnras, 477, 1004

\bibitem[{{Hunziker} {et~al.}(2018){Hunziker}, {Quanz}, {Amara}, \&
  {Meyer}}]{Hunziker2018}
{Hunziker}, S., {Quanz}, S.~P., {Amara}, A., \& {Meyer}, M.~R. 2018, \aap, 611,
  A23

\bibitem[{{Kluska} {et~al.}(2020){Kluska}, {Berger}, {Malbet}, {Lazareff},
  {Benisty}, {Le Bouquin}, {Absil}, {Baron}, {Delboulb{\'e}}, {Duvert},
  {Isella}, {Jocou}, {Juhasz}, {Kraus}, {Lachaume}, {M{\'e}nard},
  {Millan-Gabet}, {Monnier}, {Moulin}, {Perraut}, {Rochat}, {Pinte}, {Soulez},
  {Tallon}, {Thi}, {Thi{\'e}baut}, {Traub}, \& {Zins}}]{Kluska2020}
{Kluska}, J., {Berger}, J.~P., {Malbet}, F., {et~al.} 2020, \aap, 636, A116

\bibitem[{{Maire} {et~al.}(2016){Maire}, {Langlois}, {Dohlen}, {Lagrange},
  {Gratton}, {Chauvin}, {Desidera}, {Girard}, {Milli}, {Vigan}, {Zins},
  {Delorme}, {Beuzit}, {Claudi}, {Feldt}, {Mouillet}, {Puget}, {Turatto}, \&
  {Wildi}}]{Maire2016}
{Maire}, A.-L., {Langlois}, M., {Dohlen}, K., {et~al.} 2016, in Society of
  Photo-Optical Instrumentation Engineers (SPIE) Conference Series, Vol. 9908,
  Ground-based and Airborne Instrumentation for Astronomy VI, ed. C.~J.
  {Evans}, L.~{Simard}, \& H.~{Takami}, 990834

\bibitem[{{Marois} {et~al.}(2006){Marois}, {Lafreni{\`e}re}, {Doyon},
  {Macintosh}, \& {Nadeau}}]{Marois2006}
{Marois}, C., {Lafreni{\`e}re}, D., {Doyon}, R., {Macintosh}, B., \& {Nadeau},
  D. 2006, \apj, 641, 556

\bibitem[{{Marr} \& {Dong}(2022)}]{Marr22}
{Marr}, M. \& {Dong}, R. 2022, \apj, 930, 80

\bibitem[{{Mazoyer} {et~al.}(2020){Mazoyer}, {Arriaga}, {Hom},
  {Millar-Blanchaer}, {Chen}, {Wang}, {Duch{\^e}ne}, {Patience}, \&
  {Pueyo}}]{mazoyer20}
{Mazoyer}, J., {Arriaga}, P., {Hom}, J., {et~al.} 2020, in Society of
  Photo-Optical Instrumentation Engineers (SPIE) Conference Series, Vol. 11447,
  Society of Photo-Optical Instrumentation Engineers (SPIE) Conference Series,
  1144759

\bibitem[{{Meeus} {et~al.}(2001){Meeus}, {Waters}, {Bouwman}, {van den Ancker},
  {Waelkens}, \& {Malfait}}]{Meeus2001}
{Meeus}, G., {Waters}, L.~B.~F.~M., {Bouwman}, J., {et~al.} 2001, \aap, 365,
  476

\bibitem[{{M{\'e}nard} {et~al.}(2020){M{\'e}nard}, {Cuello}, {Ginski}, {van der
  Plas}, {Villenave}, {Gonzalez}, {Pinte}, {Benisty}, {Boccaletti}, {Price},
  {Boehler}, {Chripko}, {de Boer}, {Dominik}, {Garufi}, {Gratton}, {Hagelberg},
  {Henning}, {Langlois}, {Maire}, {Pinilla}, {Ruane}, {Schmid}, {van Holstein},
  {Vigan}, {Zurlo}, {Hubin}, {Pavlov}, {Rochat}, {Sauvage}, \&
  {Stadler}}]{Menard2020}
{M{\'e}nard}, F., {Cuello}, N., {Ginski}, C., {et~al.} 2020, \aap, 639, L1

\bibitem[{{Menu} {et~al.}(2015){Menu}, {van Boekel}, {Henning}, {Leinert},
  {Waelkens}, \& {Waters}}]{Menu2015}
{Menu}, J., {van Boekel}, R., {Henning}, T., {et~al.} 2015, \aap, 581, A107

\bibitem[{{Meru} {et~al.}(2017){Meru}, {Juh{\'a}sz}, {Ilee}, {Clarke},
  {Rosotti}, \& {Booth}}]{Meru2017}
{Meru}, F., {Juh{\'a}sz}, A., {Ilee}, J.~D., {et~al.} 2017, \apjl, 839, L24

\bibitem[{{Mesa} {et~al.}(2023){Mesa}, {Gratton}, {Kervella}, {Bonavita},
  {Desidera}, {D'Orazi}, {Marino}, {Zurlo}, \& {Rigliaco}}]{Mesa2023_AFLepb}
{Mesa}, D., {Gratton}, R., {Kervella}, P., {et~al.} 2023, \aap, 672, A93

\bibitem[{{Milli} {et~al.}(2018){Milli}, {Kasper}, {Bourget}, {Pannetier},
  {Mouillet}, {Sauvage}, {Reyes}, {Fusco}, {Cantalloube}, {Tristam}, {Wahhaj},
  {Beuzit}, {Girard}, {Mawet}, {Telle}, {Vigan}, \& {N'Diaye}}]{LWE_Milli_2018}
{Milli}, J., {Kasper}, M., {Bourget}, P., {et~al.} 2018, in Society of
  Photo-Optical Instrumentation Engineers (SPIE) Conference Series, Vol. 10703,
  Adaptive Optics Systems VI, ed. L.~M. {Close}, L.~{Schreiber}, \&
  D.~{Schmidt}, 107032A

\bibitem[{{Milli} {et~al.}(2012){Milli}, {Mouillet}, {Lagrange}, {Boccaletti},
  {Mawet}, {Chauvin}, \& {Bonnefoy}}]{Milli2012}
{Milli}, J., {Mouillet}, D., {Lagrange}, A.~M., {et~al.} 2012, \aap, 545, A111

\bibitem[{{Montesinos} {et~al.}(2016){Montesinos}, {Perez}, {Casassus},
  {Marino}, {Cuadra}, \& {Christiaens}}]{Montesinos2016}
{Montesinos}, M., {Perez}, S., {Casassus}, S., {et~al.} 2016, \apjl, 823, L8

\bibitem[{{Nielsen} {et~al.}(2019){Nielsen}, {De Rosa}, {Macintosh}, {Wang},
  {Ruffio}, {Chiang}, {Marley}, {Saumon}, {Savransky}, {Ammons}, {Bailey},
  {Barman}, {Blain}, {Bulger}, {Burrows}, {Chilcote}, {Cotten}, {Czekala},
  {Doyon}, {Duch{\^e}ne}, {Esposito}, {Fabrycky}, {Fitzgerald}, {Follette},
  {Fortney}, {Gerard}, {Goodsell}, {Graham}, {Greenbaum}, {Hibon}, {Hinkley},
  {Hirsch}, {Hom}, {Hung}, {Dawson}, {Ingraham}, {Kalas}, {Konopacky},
  {Larkin}, {Lee}, {Lin}, {Maire}, {Marchis}, {Marois}, {Metchev},
  {Millar-Blanchaer}, {Morzinski}, {Oppenheimer}, {Palmer}, {Patience},
  {Perrin}, {Poyneer}, {Pueyo}, {Rafikov}, {Rajan}, {Rameau}, {Rantakyr{\"o}},
  {Ren}, {Schneider}, {Sivaramakrishnan}, {Song}, {Soummer}, {Tallis},
  {Thomas}, {Ward-Duong}, \& {Wolff}}]{Nielsen2019}
{Nielsen}, E.~L., {De Rosa}, R.~J., {Macintosh}, B., {et~al.} 2019, \aj, 158,
  13

\bibitem[{{Pueyo}(2016)}]{pueyo16}
{Pueyo}, L. 2016, \apj, 824, 117

\bibitem[{{Rameau} {et~al.}(2017){Rameau}, {Follette}, {Pueyo}, {Marois},
  {Macintosh}, {Millar-Blanchaer}, {Wang}, {Vega}, {Doyon}, {Lafreni{\`e}re},
  {Nielsen}, {Bailey}, {Chilcote}, {Close}, {Esposito}, {Males}, {Metchev},
  {Morzinski}, {Ruffio}, {Wolff}, {Ammons}, {Barman}, {Bulger}, {Cotten}, {De
  Rosa}, {Duchene}, {Fitzgerald}, {Goodsell}, {Graham}, {Greenbaum}, {Hibon},
  {Hung}, {Ingraham}, {Kalas}, {Konopacky}, {Larkin}, {Maire}, {Marchis},
  {Oppenheimer}, {Palmer}, {Patience}, {Perrin}, {Poyneer}, {Rajan},
  {Rantakyr{\"o}}, {Marley}, {Savransky}, {Schneider}, {Sivaramakrishnan},
  {Song}, {Soummer}, {Thomas}, {Wallace}, {Ward-Duong}, \&
  {Wiktorowicz}}]{Rameau2017}
{Rameau}, J., {Follette}, K.~B., {Pueyo}, L., {et~al.} 2017, \aj, 153, 244

\bibitem[{{Ren} {et~al.}(2018{\natexlab{a}}){Ren}, {Dong}, {Esposito}, {Pueyo},
  {Debes}, {Poteet}, {Choquet}, {Benisty}, {Chiang}, {Grady}, {Hines},
  {Schneider}, \& {Soummer}}]{Ren2018_MWC758}
{Ren}, B., {Dong}, R., {Esposito}, T.~M., {et~al.} 2018{\natexlab{a}}, \apjl,
  857, L9

\bibitem[{{Ren} {et~al.}(2020{\natexlab{a}}){Ren}, {Dong}, {van Holstein},
  {Ruffio}, {Calvin}, {Girard}, {Benisty}, {Boccaletti}, {Esposito}, {Choquet},
  {Mawet}, {Pueyo}, {Stolker}, {Chiang}, {Boer}, {Debes}, {Garufi}, {Grady},
  {Hines}, {Maire}, {M{\'e}nard}, {Millar-Blanchaer}, {Perrin}, {Poteet}, \&
  {Schneider}}]{Ren2020_MWC758}
{Ren}, B., {Dong}, R., {van Holstein}, R.~G., {et~al.} 2020{\natexlab{a}},
  \apjl, 898, L38

\bibitem[{{Ren} {et~al.}(2020{\natexlab{b}}){Ren}, {Pueyo}, {Chen}, {Choquet},
  {Debes}, {Duch{\^e}ne}, {M{\'e}nard}, \& {Perrin}}]{Ren2020_DIsNMF}
{Ren}, B., {Pueyo}, L., {Chen}, C., {et~al.} 2020{\natexlab{b}}, \apj, 892, 74

\bibitem[{{Ren} {et~al.}(2018{\natexlab{b}}){Ren}, {Pueyo}, {Zhu}, {Debes}, \&
  {Duch{\^e}ne}}]{Ren2018_NMF}
{Ren}, B., {Pueyo}, L., {Zhu}, G.~B., {Debes}, J., \& {Duch{\^e}ne}, G.
  2018{\natexlab{b}}, \apj, 852, 104

\bibitem[{{Rich} {et~al.}(2019){Rich}, {Wisniewski}, {Currie}, {Fukagawa},
  {Grady}, {Sitko}, {Pikhartova}, {Hashimoto}, {Abe}, {Brandner}, {Brandt},
  {Carson}, {Chilcote}, {Dong}, {Feldt}, {Goto}, {Groff}, {Guyon}, {Hayano},
  {Hayashi}, {Hayashi}, {Henning}, {Hodapp}, {Ishii}, {Iye}, {Janson},
  {Jovanovic}, {Kandori}, {Kasdin}, {Knapp}, {Kudo}, {Kusakabe}, {Kuzuhara},
  {Kwon}, {Lozi}, {Martinache}, {Matsuo}, {Mayama}, {McElwain}, {Miyama},
  {Morino}, {Moro-Martin}, {Nakagawa}, {Nishimura}, {Pyo}, {Serabyn}, {Suto},
  {Russel}, {Suzuki}, {Takami}, {Takato}, {Terada}, {Thalmann}, {Turner},
  {Uyama}, {Wagner}, {Watanabe}, {Yamada}, {Takami}, {Usuda}, \&
  {Tamura}}]{Rich2019}
{Rich}, E.~A., {Wisniewski}, J.~P., {Currie}, T., {et~al.} 2019, \apj, 875, 38

\bibitem[{{Rosotti} {et~al.}(2020){Rosotti}, {Benisty}, {Juh{\'a}sz}, {Teague},
  {Clarke}, {Dominik}, {Dullemond}, {Klaassen}, {Matr{\`a}}, \&
  {Stolker}}]{Rosotti2020}
{Rosotti}, G.~P., {Benisty}, M., {Juh{\'a}sz}, A., {et~al.} 2020, \mnras, 491,
  1335

\bibitem[{{Safonov} {et~al.}(2022){Safonov}, {Strakhov}, {Goliguzova}, \&
  {Voziakova}}]{Safonov2022}
{Safonov}, B.~S., {Strakhov}, I.~A., {Goliguzova}, M.~V., \& {Voziakova}, O.~V.
  2022, \aj, 163, 31

\bibitem[{{Shuai} {et~al.}(2022){Shuai}, {Ren}, {Dong}, {Zhou}, {Pueyo}, {De
  Rosa}, {Fang}, \& {Mawet}}]{Shuai2022}
{Shuai}, L., {Ren}, B.~B., {Dong}, R., {et~al.} 2022, \apjs, 263, 31

\bibitem[{{Soummer} {et~al.}(2012){Soummer}, {Pueyo}, \&
  {Larkin}}]{Soummer2012}
{Soummer}, R., {Pueyo}, L., \& {Larkin}, J. 2012, \apjl, 755, L28

\bibitem[{{van der Plas} {et~al.}(2019){van der Plas}, {M{\'e}nard},
  {Gonzalez}, {Perez}, {Rodet}, {Pinte}, {Cieza}, {Casassus}, \&
  {Benisty}}]{vanderPlas2019}
{van der Plas}, G., {M{\'e}nard}, F., {Gonzalez}, J.~F., {et~al.} 2019, \aap,
  624, A33

\bibitem[{{Vigan}(2020)}]{SPHERE_pipeline_Vigan2020}
{Vigan}, A. 2020, {vlt-sphere: Automatic VLT/SPHERE data reduction and
  analysis}

\bibitem[{{Vigan} {et~al.}(2021){Vigan}, {Fontanive}, {Meyer}, {Biller},
  {Bonavita}, {Feldt}, {Desidera}, {Marleau}, {Emsenhuber}, {Galicher}, {Rice},
  {Forgan}, {Mordasini}, {Gratton}, {Le Coroller}, {Maire}, {Cantalloube},
  {Chauvin}, {Cheetham}, {Hagelberg}, {Lagrange}, {Langlois}, {Bonnefoy},
  {Beuzit}, {Boccaletti}, {D'Orazi}, {Delorme}, {Dominik}, {Henning}, {Janson},
  {Lagadec}, {Lazzoni}, {Ligi}, {Menard}, {Mesa}, {Messina}, {Moutou},
  {M{\"u}ller}, {Perrot}, {Samland}, {Schmid}, {Schmidt}, {Sissa}, {Turatto},
  {Udry}, {Zurlo}, {Abe}, {Antichi}, {Asensio-Torres}, {Baruffolo}, {Baudoz},
  {Baudrand}, {Bazzon}, {Blanchard}, {Bohn}, {Brown Sevilla}, {Carbillet},
  {Carle}, {Cascone}, {Charton}, {Claudi}, {Costille}, {De Caprio},
  {Delboulb{\'e}}, {Dohlen}, {Engler}, {Fantinel}, {Feautrier}, {Fusco},
  {Gigan}, {Girard}, {Giro}, {Gisler}, {Gluck}, {Gry}, {Hubin}, {Hugot},
  {Jaquet}, {Kasper}, {Le Mignant}, {Llored}, {Madec}, {Magnard}, {Martinez},
  {Maurel}, {M{\"o}ller-Nilsson}, {Mouillet}, {Moulin}, {Orign{\'e}}, {Pavlov},
  {Perret}, {Petit}, {Pragt}, {Puget}, {Rabou}, {Ramos}, {Rickman}, {Rigal},
  {Rochat}, {Roelfsema}, {Rousset}, {Roux}, {Salasnich}, {Sauvage}, {Sevin},
  {Soenke}, {Stadler}, {Suarez}, {Wahhaj}, {Weber}, \& {Wildi}}]{Vigan2021}
{Vigan}, A., {Fontanive}, C., {Meyer}, M., {et~al.} 2021, \aap, 651, A72

\bibitem[{{Vigan} {et~al.}(2010){Vigan}, {Moutou}, {Langlois}, {Allard},
  {Boccaletti}, {Carbillet}, {Mouillet}, \& {Smith}}]{Vigan2010}
{Vigan}, A., {Moutou}, C., {Langlois}, M., {et~al.} 2010, \mnras, 407, 71

\bibitem[{{Vioque} {et~al.}(2018){Vioque}, {Oudmaijer}, {Baines},
  {Mendigut{\'\i}a}, \& {P{\'e}rez-Mart{\'\i}nez}}]{Vioque2018}
{Vioque}, M., {Oudmaijer}, R.~D., {Baines}, D., {Mendigut{\'\i}a}, I., \&
  {P{\'e}rez-Mart{\'\i}nez}, R. 2018, \aap, 620, A128

\bibitem[{{Wagner} {et~al.}(2018){Wagner}, {Dong}, {Sheehan}, {Apai}, {Kasper},
  {McClure}, {Morzinski}, {Close}, {Males}, {Hinz}, {Quanz}, \&
  {Fung}}]{Wagner2018}
{Wagner}, K., {Dong}, R., {Sheehan}, P., {et~al.} 2018, \apj, 854, 130

\bibitem[{{Xie} {et~al.}(2022){Xie}, {Choquet}, {Vigan}, {Cantalloube},
  {Benisty}, {Boccaletti}, {Bonnefoy}, {Desgrange}, {Garufi}, {Girard},
  {Hagelberg}, {Janson}, {Kenworthy}, {Lagrange}, {Langlois}, {Menard}, \&
  {Zurlo}}]{xie22}
{Xie}, C., {Choquet}, E., {Vigan}, A., {et~al.} 2022, \aap, 666, A32

\bibitem[{{Xie} {et~al.}(2020){Xie}, {Haffert}, {de Boer}, {Kenworthy},
  {Brinchmann}, {Girard}, {Snellen}, \& {Keller}}]{Xie2020}
{Xie}, C., {Haffert}, S.~Y., {de Boer}, J., {et~al.} 2020, \aap, 644, A149

\bibitem[{{Xie} {et~al.}(2021){Xie}, {Ren}, {Dong}, {Pueyo}, {Ruffio}, {Fang},
  {Mawet}, \& {Stolker}}]{Xie_Chengyan2021}
{Xie}, C., {Ren}, B., {Dong}, R., {et~al.} 2021, \apjl, 906, L9

\bibitem[{{Zhu} \& {Zhang}(2022)}]{Zhu2022}
{Zhu}, Z. \& {Zhang}, R.~M. 2022, \mnras, 510, 3986

\bibitem[{{Zurlo} {et~al.}(2020){Zurlo}, {Cugno}, {Montesinos}, {Perez},
  {Canovas}, {Casassus}, {Christiaens}, {Cieza}, \& {Huelamo}}]{Zurlo2020}
{Zurlo}, A., {Cugno}, G., {Montesinos}, M., {et~al.} 2020, \aap, 633, A119

\end{thebibliography}

\begin{appendix}

\section{SPHERE/IRDIS observations and data reduction}
\label{appendix:observation_log}

SPHERE/IRDIS has performed multiple K-band observations of the HD\,100453 system in 2015, 2016, and 2019. However, the observing conditions were extremely bad for two observations in 2016 (see Table~\ref{table:obs_log}), resulting in the strong low wind effect \citep{LWE_Milli_2018} that affects the disk morphology. Therefore, we excluded the 2016 data from the dynamical analysis. Although IRDIS also has an H-band observation in 2016, it is essential to perform the motion measurement based on observations in the same wavelength  because different wavelengths trace dust grains with different sizes, which are possibly located at different places in the disk. So the K-band observations in 2015 and 2019 are the only available data for HD\,100453 to perform the dynamical analysis.

All the observations used the apodized pupil Lyot coronagraph in its $\texttt{N\_ALC\_YJH\_S}$ configuration \citep{Carbillet2011, Guerri2011}, with a mask diameter of 185~mas and a pixel scale of 12.25 mas. IRDIS in its dual-band imaging \citep[DBI;][]{Vigan2010} mode produces simultaneous images at two nearby wavelengths (K1: 2.110~$\mu$m and K2: 2.251~$\mu$m). While K-band observation is affected by the thermal background emission, all of the observations did not contain sky background calibrations to calibrate it. We therefore used DIsNMF to model and generate the synthetic sky background images based on all the available sky background images in the SPHERE archive, similar to the technique described in \cite{Hunziker2018}. Each observation was then processed using the $\texttt{vlt-sphere}$\footnote{\url{https://github.com/avigan/SPHERE}, version 1.4.2} pipeline \citep{SPHERE_pipeline_Vigan2020} to correct for the sky background based on our NMF sky model, flat field, and bad pixels. Then the pipeline generated calibrated and roughly aligned (offsets\textless1 pix) data cube for each observation. 

In the K1 band, our DIsNMF approach can successfully model and then remove the thermal background. However, we did not cleanly remove all the thermal background emissions in K2  because of the stronger background variation at longer wavelengths (i.e., K2) and limited calibration images in the SPHERE archive for modeling. To avoid potential influence from the residual of the background in K2, we only use K1 data to measure the spiral motion in this work.
\FloatBarrier

\subsection{Image alignment}
\label{appendix:image_alignment}

To determine the star center behind the coronagraph, SPHERE generates satellite spots on coronagraph images by introducing a 2D periodic modulation on the high-order deformable mirror \citep{Beuzit2019}, obtaining   star center images. SPHERE usually uses satellite spots in the first and last images of an observation to locate the star center behind the coronagraph. During the entire science observation of HD\,100453, SPHERE relies on the differential tip-tilt sensor control to maintain the star at the same position behind the coronagraphic mask. However, the differential tip-tilt sensor loop runs at 1 Hz, so some residual jitter of the images can occur at a faster rate, therefore inducing a small shift (typically \textless1 pix; \citealp{xie22}). 

The misalignment of images within each epoch of observation and between different observations has a direct impact on motion measurement. We performed the fine alignment for all the science images in two steps, the alignment of star center images from different epochs, and the alignment of science images to their corresponding star center images. We found that the star center images from three epochs were properly aligned by $\texttt{vlt-sphere}$ with offsets less than 0.05 pix by measuring the intersection of four satellite spots. So no additional alignment was required for star center images from different epochs.

To align science images within each epoch of observation, we used the position of the companion. 
The high signal-to-noise ratio (S/N$\sim$588) of the companion enables the fine alignment of the science images after the pre-processing of the $\texttt{vlt-sphere}$ pipeline. In the star center image, the positions of the companion and the primary star are known. 
Since we knew the field rotation between the star center image and a given science image, we rotated the star center image to create a reference position of the companion when the given science image was taken. The positions of the reference companion and the companion in the science image were then determined by 2D Gaussian fitting because the companion has no known disk. The derived position offset is the offset between the star center image and a given science image. Once we obtained all the offsets, we performed image shifts to create a finely aligned and calibrated science data cube for each epoch.  

We examined the offset of the companion position in two observations of 2019 after the companion alignment and post-processing described in Appendix~\ref{appendix:RDI-DIsNMF}. The obtained offset is less than 0.06  pixels, which is the residual offset in the star center images. In summary, our image alignment can accurately align all the science images to a common reference, thus avoiding the false positive caused by the instrument in the motion measurement. 

\begin{table}
\centering
\setlength{\tabcolsep}{2pt}
\caption{SPHERE/IRDIS observations of the HD\,100453 system \label{table:obs_log}}
\begin{tabular}{l rcc r}   \hline\hline
Prog. ID & Date & Band  & $n_{\rm DIT} \times t_{\rm DIT}$\tablenotemark{a} & Wind speed\\
 &   &  &  (s) &  (m~s$^{-1}$) \\ \hline
095.C-0389(A)  &   2015 Apr 10          &       K12     &  $176 \times 8$ &  $8.54 \pm 0.36$       \\
095.C-0389(A)\tablenotemark{b} &   2016 Jan 16          &       K12     &  $176 \times 8$ &       $1.56 \pm 0.78$          \\
095.C-0389(A)\tablenotemark{b}  &   2016 Jan 23         &       K12     &  $176 \times 8$ &       $4.88 \pm 0.19$          \\
0103.C-0847(A)  &   2019 Apr 07         &       K12     &  $42 \times 32$ &       $3.64 \pm 0.19$          \\
0103.C-0847(A)  &   2019 Apr 08         &       K12     &  $26 \times 32$ &       $4.40 \pm 0.33$    \\ \hline
\end{tabular}

\begin{flushleft}

{\small \textbf{Notes}: 

$^a${$n_{\rm DIT}$ is the number of image frames and $t_{\rm DIT}$ is exposure time per image frame.}
$^b${We do not use two observations in 2016 in this work because of their bad observing conditions that strongly affect the disk morphology. Before new coatings on the telescope spiders in 2017, SPHERE was affected by the low wind effect for wind speeds  lower than 5~m~s$^{-1}$ \citep{LWE_Milli_2018}. Such a low wind effect results in strong stellar-light leakage around the IWA and strong spider patterns (i.e., diffraction spikes). As a result, the spider patterns strongly alter the disk morphology, which prevents us from correctly measuring the locations of spiral arms.}

}
\end{flushleft}

\end{table}

\subsection{Bad frame exclusion}
Bad frame exclusion does not alter the true morphology of the disk. The aim is to have a higher S/N of the disk by increasing the disk signal (i.e., including more images) and reducing the distortion from bad frames (i.e., reducing instrumental residuals). Failed adaptive optics (AO) corrections cause the host star outside the coronagraph, which should be excluded. Bad AO corrections result in strong stellar-light leakage around the inner working angle (IWA) and clear spider patterns. Such spider patterns were not easily  removed by RDI, and hence left strong residuals in the disk image that altered the disk morphology. Therefore, we also excluded the images with strong spider patterns, mainly in the 2015 observation. In total, we excluded about 77\%, 14\%, and 19\% of the science images for observations in 2015, and on 07 and 08 April 2019, respectively. Thanks to the high disk surface brightness, we had enough disk signal to perform the dynamical analyses after the bad frame exclusion. 

\section{Stellar emission subtraction}

\label{appendix:RDI-DIsNMF}

We removed the stellar contribution using RDI-DIsNMF to map the disk. Although other techniques based on angular differential imaging \citep[ADI;][]{Marois2006} are common approaches to remove stellar point spread function (PSF), it usually produces  nonphysical artifacts when applied to disks \citep{Milli2012}. For example, ADI has the self-subtraction effect that lowers the throughput of the disk and more importantly alters the disk morphology. Because ADI builds the PSF reference based on the science data itself, it may contain some of the astrophysical signals in the PSF model. Unlike ADI, RDI builds the PSF references from the companion-free and disk-free reference images. Therefore, RDI naturally avoided the self-subtraction effect. A commonly used PSF reconstruction technique is principal component analysis \citep[PCA;][]{Soummer2012}. In data processing, however, PCA removes the mean of the image, and thus creates nonphysical negative regions around a strong astrophysical signal (i.e., a bright disk), calling for forward modeling to properly recover these signals \citep[e.g.,][]{pueyo16, mazoyer20}. In comparison, NMF does not remove the mean of the image \citep{Ren2018_NMF}, which can thus avoid creating negative regions around bright sources in data pre- and post-processing. Furthermore, the recent development of the NMF algorithm by \citet{Ren2020_DIsNMF} introduces the data imputation concept, which ignores the disk region to avoid the overfitting problem when reconstructing the PSF model.

\subsection{Initial PSF Selection}
We followed the method described in \citet{xie22}~to perform RDI. The key step in RDI is assembling a proper PSF reference library. We created the master PSF reference library by using all the public archival data in K1 taken with IRDIS under the same coronagraphic settings. 
The pre-processing of all the archival data was performed using $\texttt{vlt-sphere}$ pipeline. 
Bad reference stars that contain astrophysical sources were excluded by the visual inspection of the residual images after the reductions of ADI and then RDI. After assembling a master reference library for K1 data, we down-selected 200 best-matched reference images for each science image in each observation of HD\,100453. For each observation, we combined the down-selected reference images and formed a single library with nonredundant references. As a result, the final sizes of reference libraries were 2802, 2646, and 1816 for observations in April 2015, and  07 and 08 April  2019, respectively.

To remove the stellar PSF we used NMF to create components of the PSF model from the reference library. After that, the PSF model was then reconstructed for each science image using the data imputation in DIsNMF after masking the source region (i.e., the disk and the companion). Finally, the residual science cube after the PSF subtraction was derotated and mean combined to form the residual image in K1.

\subsection{Final PSF Selection}
\begin{figure}[h!]
\centering
        \includegraphics[width=0.49\textwidth]{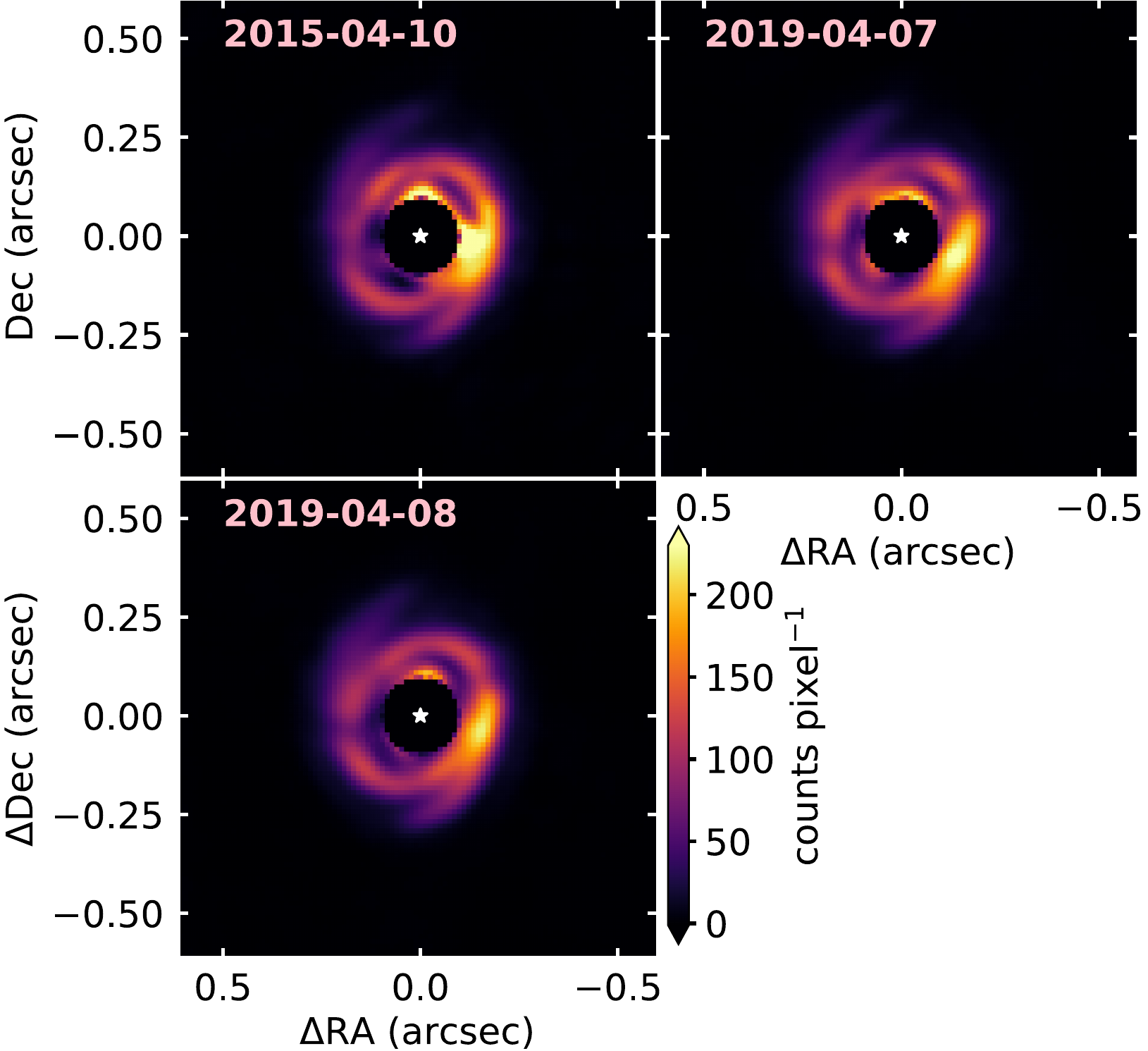}
    \caption{Multi-epoch detection of the spiral disk around HD\,100453. The disk is imaged in K1 and reduced with RDI-DIsNMF to remove the direct starlight. The white star shows the position of the host star.}
    \label{appendix_fig1}
\end{figure}
\FloatBarrier
The disk of HD\,100453 is bright enough to affect the down-selection of the reference images using the mean square error described in \citet{xie22}. Consequently, the selected reference images tend to have certain levels of wind-driven halo \citep{Cantalloube2020} that mimic bright and extended disk features. As a result, poorly matched PSF references lead to oversubtraction that sightly affects the disk morphology. Better-matched PSF references can be selected if the disk contribution is significantly reduced. 

\begin{figure}[h!]
\centering

        \includegraphics[width=0.49\textwidth]{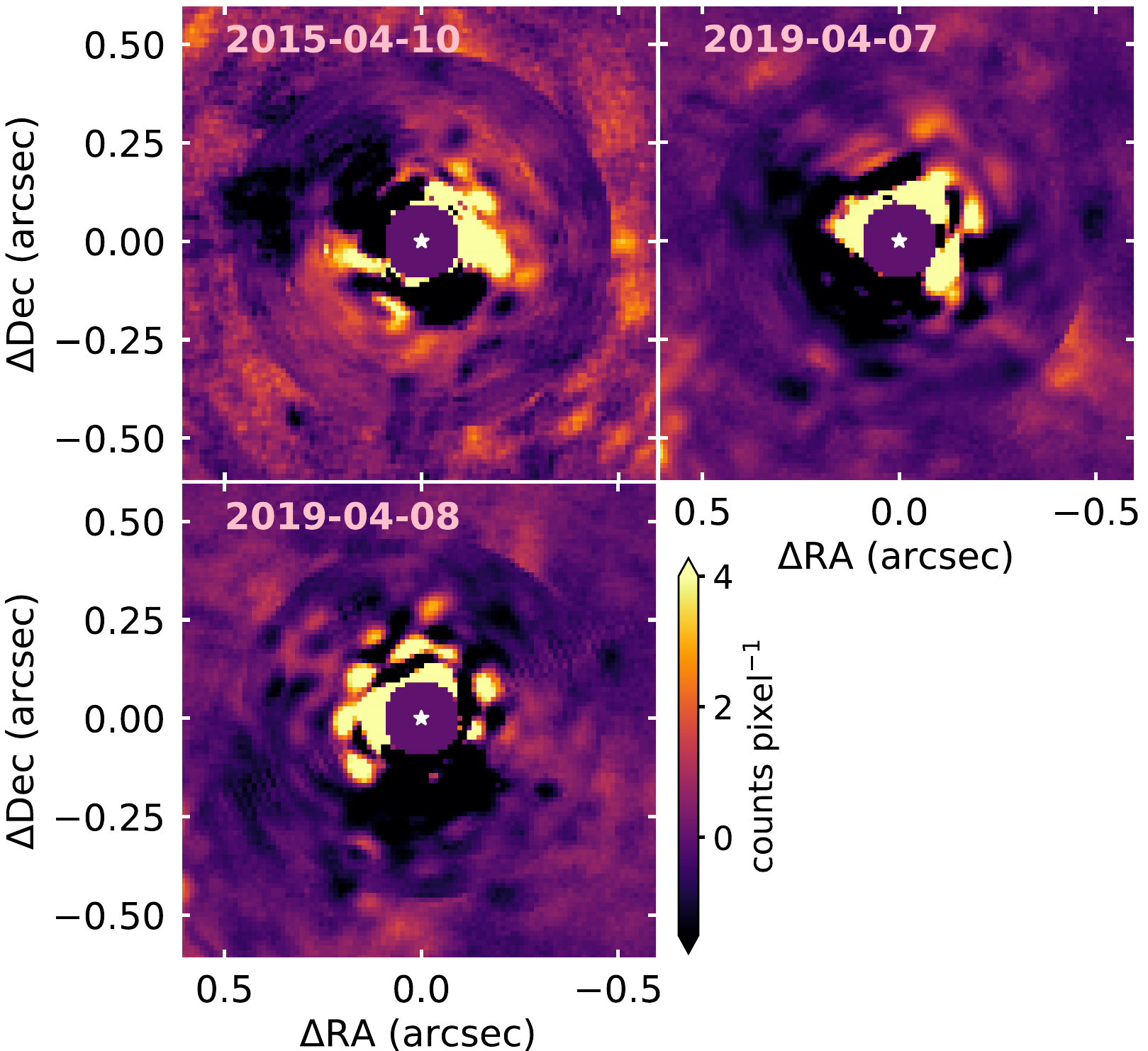}
    \caption{Residual images of disk-removed science data after the reduction of RDI-DIsNMF. No disk signal is left over, indicating a good recovery of disk flux. The white star shows the position of the host star. }
    \label{appendix_fig2_residual_disk_free}
\end{figure}

For the final PSF selection, we adopted an iterative process in which we first removed the disk obtained with RDI-DIsNMF from the science data, then re-selected a better matching library of PSF references with the disk-removed science images. Only the reference library that is selected for PSF modeling was updated during each iteration. The original science images remain unchanged in each RDI-DIsNMF subtraction.
For the HD~100453 exposures in this study, the disk-removed science images are sufficiently clear of disk signals to converge on matching PSF references after a maximum of five iterations. In Fig.~\ref{appendix_fig2_residual_disk_free} we show the residual images of disk-removed science data after the reduction of RDI-DIsNMF. No disk signal is left over, indicating a good recovery of disk flux. Because we directly subtracted the disk image to obtain the disk-removed science images, the noise pattern was changed in the disk region (\textless 0.45\arcsec).
We performed a final RDI-DIsNMF subtraction to obtain the disk image for pattern motion analysis (see Fig.~\ref{appendix_fig1}). Throughout the paper, the observation of the disk on 08 April  2019~is only used for uncertainty estimation (see Appendix~\ref{appendix:uncertainty_estimation}) because the integration time of the second epoch observation in 2019 is shorter than the first one (see Table~\ref{table:obs_log}). 

\FloatBarrier

\FloatBarrier

\section{Determining the location of the spiral arm}
\label{appendix:arm_location_fit}
To determine the local maxima of the spiral arm in polar coordinates, we performed two Gaussian profile fits with an additional constant, totaling seven free parameters. By doing so, we can simultaneously account for the existence of a ring-like structure near the IWA and that of a spiral. The additional constant was adopted in the model to account for the overall disk emission. 

The regions with radii of \textless 14 pixels and \textgreater 35 pixels were masked out to avoid the noisy regions close to the IWA (8 pixels) and those without disk emissions, respectively. In each azimuthal angle (1$^{\circ}$ step) in the polar images  we performed a fitting to obtain the location of the spiral arm and presented it in Figs.~\ref{check_fit_motion_NE} and \ref{appendix_check_NMF_center_fit_NE_ERROR}. We visually inspected all fitting results to ensure the correctness of the fitting (i.e., that the data are reasonably represented by the model).

\section{Uncertainty estimation}
\label{appendix:uncertainty_estimation}

\FloatBarrier
Precise motion measurements require accurate recovery of disk morphology. While RDI-DIsNMF can avoid the self-subtraction effect and it is theoretically expected to mitigate the overfitting of stellar PSFs, it may still slightly alter the disk morphology during PSF modeling since we cannot guarantee a perfect match of the speckles between a target image and its corresponding selected PSF references. 

To account for this potential PSF mismatch effect, it is necessary to introduce additional uncertainty in spiral motion measurements. With the two K1 observations in 2019 obtained on different nights, the temporal separation is too small (1 day apart) to obtain spiral motion measurement. However, these two observations in 2019 are ideal in offering a unique opportunity to examine the unknown uncertainties in our dynamical analysis, especially in quantifying the change in disk morphology, and thus its impact on spiral motion caused by RDI-DIsNMF. 

\begin{figure}[htb!]
\centering
        \includegraphics[width=0.47\textwidth]{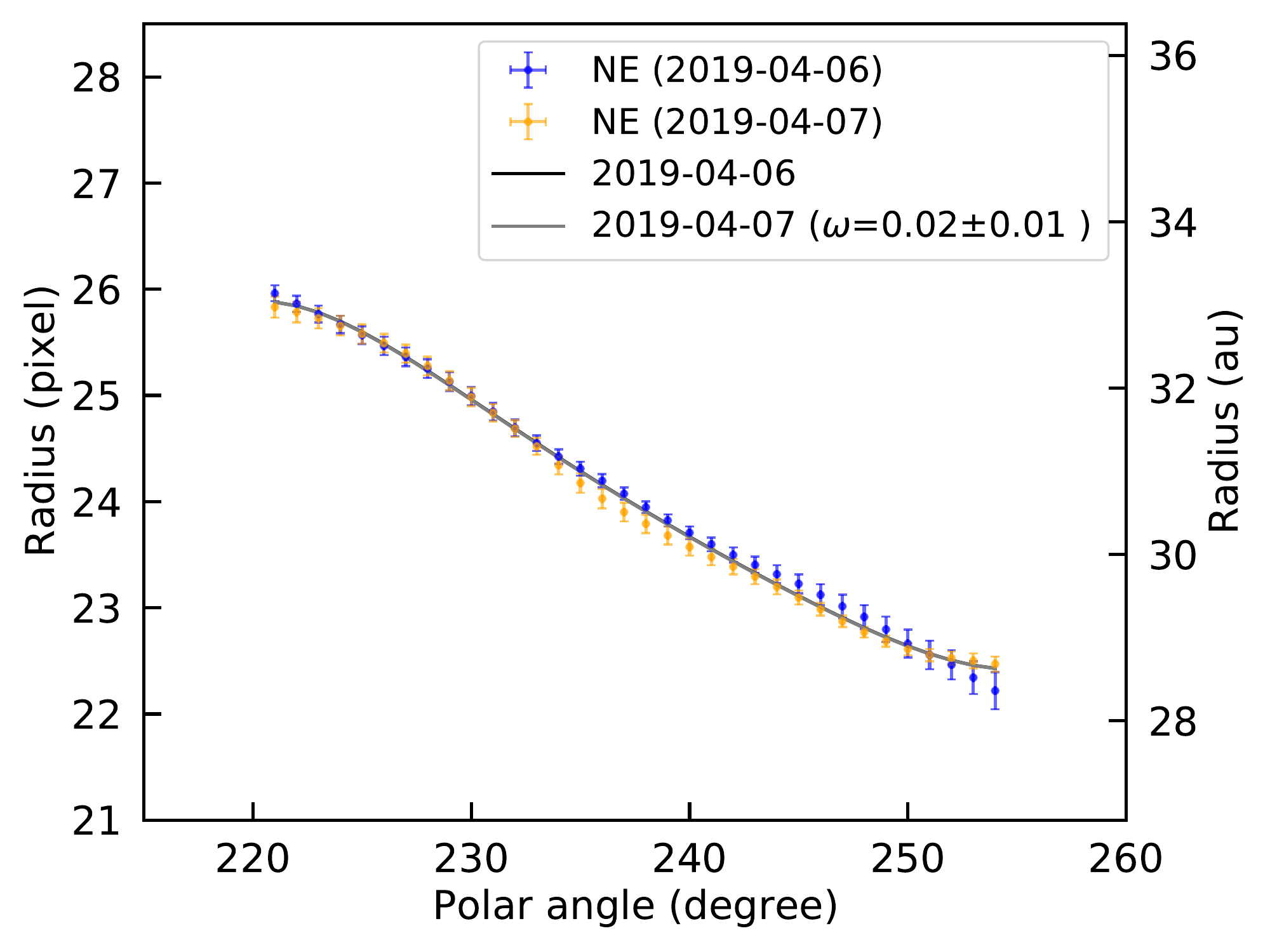}
    \caption{Peak locations of spiral arm S1 in polar coordinates after the correction for the viewing geometry. Solid curves represent the best-fit model spiral for the peak locations (dot) between two epoch observations in 2019, assuming the companion-driven scenario. Two curves (black and gray) are overlapped. The derived angular velocities of spiral pattern motion are $0\fdg02$~yr$^{-1}$, which indicates the uncertainty caused by PSF subtraction instead of the real motion.
    }
    \label{appendix_check_NMF_center_fit_NE_ERROR}
\end{figure}

By measuring the motion of the spiral arms in two observations in 2019, we can estimate the motion caused by our post-processing method instead of the real spiral motion.  Given the time span of $\sim$1 day, the real spiral motion is $\sim$0 degree. We performed the identical motion measurement procedure as in Sect.~\ref{sec:Analysis}, and obtained a motion of $0\fdg02$ for the S1 arm in the two 2019 observations  shown in Fig.~\ref{appendix_check_NMF_center_fit_NE_ERROR}. Nevertheless, it is possible that the selected PSFs do not necessarily return such uncertainties for all the epochs studied here. Therefore, we conservatively consider an additional uncertainty of $\sigma_{\text{RDI} } = 0\fdg1$ for the RDI-DIsNMF method.

In our analysis, the total uncertainty in our pattern speed analysis is 
\begin{equation}\label{eq:uncertainty}
\sigma =\sqrt{\sigma^{2}_{\text{fit} } +\sigma^{2}_{\text{north} } +\sigma^{2}_{\text{RDI} } } \  t^{-1},
\end{equation}
where $\sigma_{\rm fit}$, $\sigma_{\rm north}$, and $\sigma_{\rm RDI}$ are uncertainties caused by the measurement of the spiral locations, true north uncertainty of SPHERE, and our post-processing method, respectively. The time span between two epochs is represented by $t$, which is 4.0 years. 

The uncertainty of the spiral S1 locations returns a fitting uncertainty ($\sigma_{\rm fit}$) of $0\fdg192$. We adopt the true north uncertainty of SPHERE to be $0\fdg08$ in all epochs \citep{Maire2016}. The uncertainty caused by the post-processing method is estimated to be $0\fdg1$ per epoch using the 2019 observations. For the motion measurement on two epochs, $\sigma_{\rm north}$ is $\sqrt{2 \times (0\fdg08)^{2}} = 0\fdg113$ and $\sigma_{\rm RDI}$ is $\sqrt{2 \times (0\fdg1)^{2}} = 0\fdg142$. 
Based on Equation~\eqref{eq:uncertainty}, the final 1$\sigma$ uncertainty on the pattern speed of the S1 arm is $0\fdg066$~yr$^{-1}$.

\section{Effect of  disk flaring on spiral motion measurement}
\begin{figure*}[htb!]
\centering
        \includegraphics[width=0.9\textwidth]{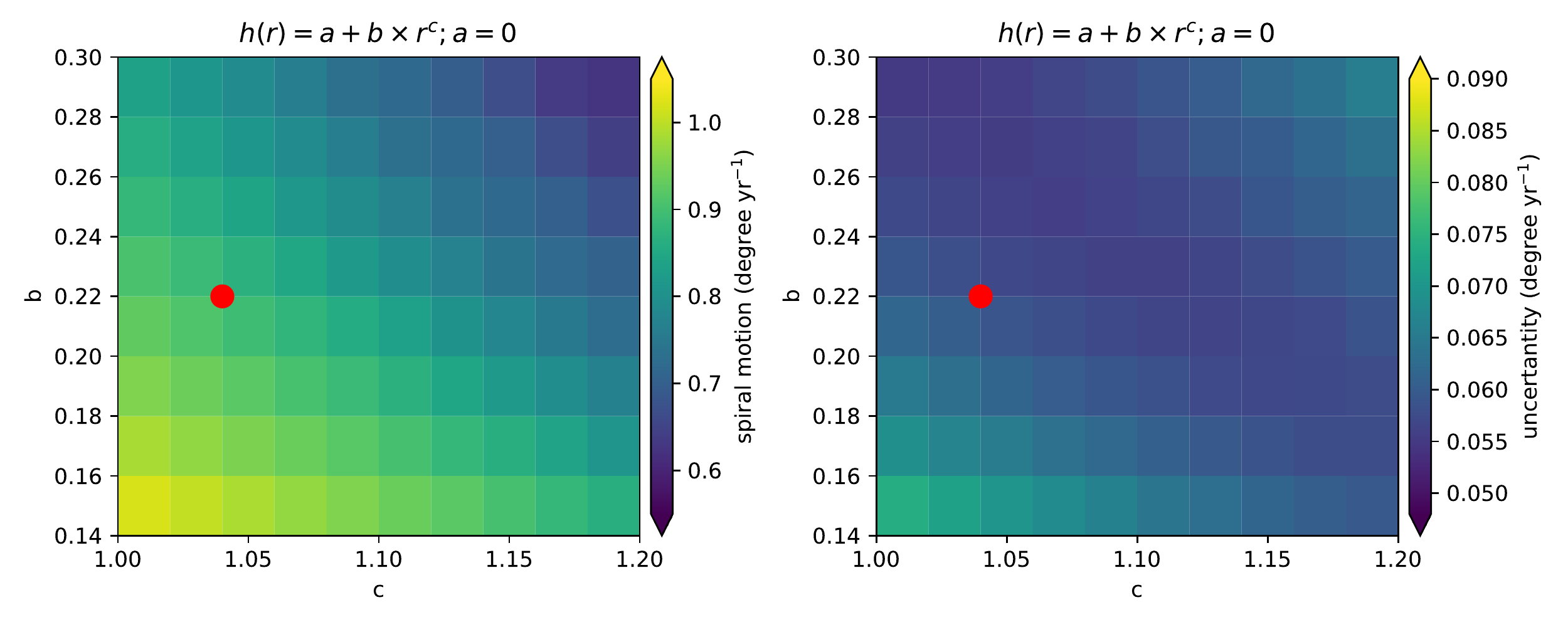}
    \caption{Effect of disk flaring on spiral motion measurement. Different disk flaring will change the spiral location determination in the deprojection, which affects the spiral motion measurement. \textit{Left panel:} Spiral motions derived based on different disk flaring. \textit{Right panel:} Corresponding uncertainties obtained using Equation~\ref{eq:uncertainty}. The red dot represents the velocity of the spiral motion and its corresponding uncertainty based on the best-fit disk model ($h = 0.22 \times r^{1.04}$) from \cite{Benisty2017}.
    }
    \label{appendix_check_disk_scale_height}
\end{figure*}
\FloatBarrier
\label{appendix:disk_scale_height}
Throughout the paper, we present the velocity of the spiral motion based on the best-fit model ($h = 0.22 \times r^{1.04}$) from \cite{Benisty2017} to correct for the disk flaring in the deprojection. The deprojection will affect the spiral location determination, and subsequently the spiral motion measurement. To study the effect of the disk flaring on spiral motion measurement, we adopted a reasonable range of parameters for correcting the disk flaring and performed new motion measurements, as described in Sect.~\ref{subsec:motion_measurement_S1}. The effect of the disk flaring on spiral motion measurement is shown in Fig.~\ref{appendix_check_disk_scale_height}.

In the case of HD\,100453, the velocity of the spiral motion decreases with the increase in the disk flaring, ranging from $1\fdg0$~yr$^{-1}$ to $0\fdg5$~yr$^{-1}$. This  velocity range still favors the companion-driven scenario and disfavors the GI scenario.  Given  that different disk flarings only have a minor impact on the spiral motion and do not change our conclusion, we only present the motion measurement based on the best-fit model of disk flaring from \cite{Benisty2017}.

\section{Companion orbital fitting}
\FloatBarrier

\begin{figure}[htb!]
\centering
        \includegraphics[width=0.42\textwidth]{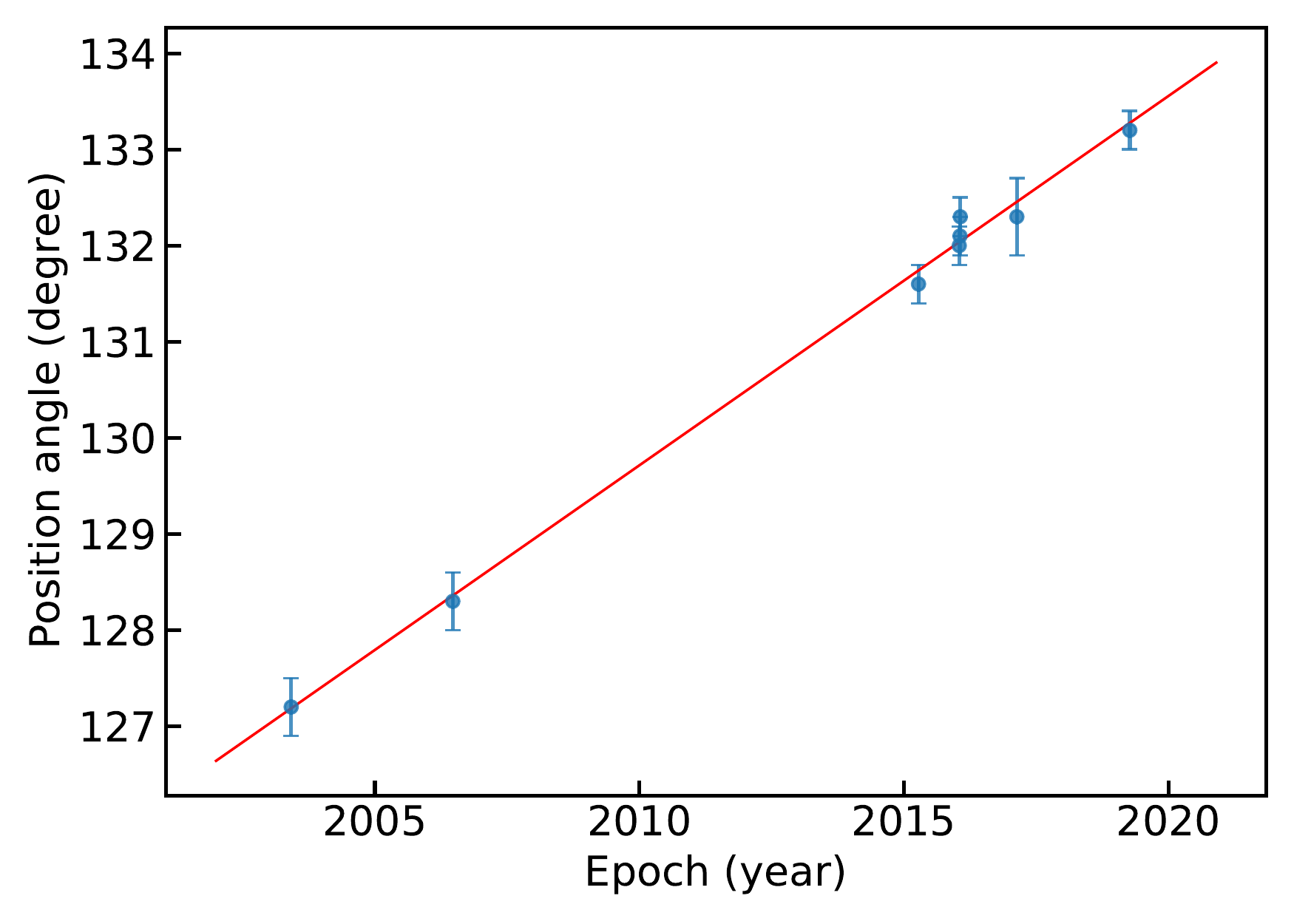}
    \caption{Linear fit to the position angles of HD\,100453\,B as the function of epochs. The slope corresponds to the measured angular velocity of the companion in the sky plane, which is $0\fdg384\pm0\fdg019$~yr$^{-1}$. The positive direction is counterclockwise, the same as the direction of the disk rotation.}
    \label{fig:astrometry_PA}
\end{figure}

\label{appendix:linear_fit_orbital_motion}
We performed a linear fit to the position angles of HD\,100453\,B from 2003 to 2019. The astrometric data is listed in Table~\ref{table:astrometry}, which was adopted from \cite{Collins2009}, \cite{Wagner2018}, and \cite{Gonzalez2020}. The orbital period of the companion is about 800 years, which is significantly longer than the time span of our astrometric data. For the ${\sim}20$-year temporal separation studied here, the linear fit is therefore sufficient in obtaining the angular velocity of the companion. Figure~\ref{fig:astrometry_PA} shows the position angles of HD\,100453\,B and our linear fitting result. The slope corresponds to the measured angular velocity of the companion in the sky plane, which is $0\fdg384\pm0\fdg019$~yr$^{-1}$ in the counterclockwise direction. Using only two astrometry data sets in 2015 and 2019, we obtained an angular velocity of $0\fdg40\pm0\fdg07$~yr$^{-1}$ in the sky plane, which validates the choice of a linear fit.

\FloatBarrier

\begin{table}
\centering
\setlength{\tabcolsep}{3pt}
\caption{Astrometric data\label{table:astrometry}}
\begin{tabular}{l rcc r}    \hline\hline
Date & Instrument & Separation & Position angle & Ref. \\ \hline
2003 Jun 02     &       NACO    &       $1\farcs049 \pm 0\farcs007$     &       $127\fdg2 \pm 0\fdg3$ & (a, b)     \\
2006 Jun 22     &       NACO    &       $1\farcs042 \pm 0\farcs005$     &       $128\fdg3 \pm 0\fdg3$ & (a, b)     \\
2015 Apr 10     &       SPHERE  &       $1\farcs047 \pm 0\farcs003$     &       $131\fdg6 \pm 0\fdg2$ & (c)        \\
2016 Jan 16     &       SPHERE  &       $1\farcs045 \pm 0\farcs003$     &       $132\fdg0 \pm 0\fdg2$ & (d)        \\
2016 Jan 21     &       SPHERE  &       $1\farcs049 \pm 0\farcs003$     &       $132\fdg1 \pm 0\fdg2$ & (d)        \\
2016 Jan 23     &       SPHERE  &       $1\farcs048 \pm 0\farcs002$     &       $132\fdg3 \pm 0\fdg2$ & (d)        \\
2017 Feb 17     &       MagAO   &       $1\farcs056 \pm 0\farcs005$     &       $132\fdg3 \pm 0\fdg4$ & (c)        \\
2019 Apr 07     &       SPHERE  &       $1\farcs046 \pm 0\farcs003$     &       $133\fdg2 \pm 0\fdg2$ & (d)        \\ \hline
\end{tabular}

\begin{flushleft}
{\small \textbf{References}: (a) \cite{Collins2009}; (b) \cite{Chauvin2010}; (c) \cite{Wagner2018}; and (d) \cite{Gonzalez2020}}
\end{flushleft}
\end{table}
\FloatBarrier

\FloatBarrier

\section{Companion orbital motion}
\label{appendix:pdf_companion_orbit}
The companion orbit can be described by the orbital elements. The radial separation ($r$) between the companion and the primary star is 
\begin{equation}\label{eq:r}
r\left( \nu \right)  =\frac{a\left( 1-e^{2}\right)  }{1+e\cos \nu } ,
\end{equation}
where $\nu$, $a$, and $e$ are the true anomaly, semimajor axis, and eccentricity, respectively.

We introduced $u$ to represent the angle between the radial direction of the companion ($r$) and the intersection between the orbit and the sky planes. So the angle $u$ is  
\begin{equation}\label{eq:u}
u=\pi -\left( \nu +\omega \right),
\end{equation}
where $\omega$ is the argument of periastron.

The projected radial separation of the companion in the sky plane ($r_{\rm proj}$) is
\begin{equation}\label{eq:r_proj}
r_{\text{proj} }\left( \nu \right)  \  =\  \frac{a\left( 1-e^{2}\right)  }{1+e\cos \nu } \sqrt{{}\cos^{2} u+\sin^{2}u\cos^{2} i} ,
\end{equation}
where $i$ is the inclination between the orbital plane and the sky plane. Equation~\eqref{eq:r_proj} shows the projection of the radial separation in Equation~\eqref{eq:r} onto the sky plane. For a given combination of orbital elements and the projected radial separation ($r_{\rm proj}$), we can obtain their corresponding true anomaly ($\nu$) by solving Equation~\eqref{eq:r_proj} numerically.

The specific angular momentum
\begin{equation}\label{eq:vector_h}
\vec{h} =\vec{r} \  \times \dot{\vec{r}  },
\end{equation}
or
\begin{equation}\label{eq:h}
h=rV_{\bot },
\end{equation}
where $V$ is the velocity of the companion in the orbit and the symbol $\bot$ donates the direction that is perpendicular to the outward radial from the primary to the companion. The orbit equation that defines the separation between the primary and the companion is
\begin{equation}\label{eq:orbit_equation}
r=\frac{h^{2}}{\mu \left( 1+e\cos \nu \right)  },
\end{equation}
where $h$ is the specific angular momentum. Substituting Equation~\eqref{eq:r} into Equation~\eqref{eq:orbit_equation}, we obtain
\begin{equation}\label{eq:hh}
h=\sqrt{\mu a\left( 1-e^{2}\right)  }.
\end{equation}

The definition of angular velocity is
\begin{equation}\label{eq:definition_angular_velocity}
\dot{u} =\frac{V_{\bot }}{r},
\end{equation}
from which we can obtain the angular velocity of the companion at a given position in the orbital plane,
\begin{equation}\label{eq:angular_velocity}
\dot{u} =\sqrt{\frac{\mu \left( 1+e\cos \nu \right)  }{r^{3}} } ,
\end{equation}
where $\mu$ is the gravitational parameter. In our case, the gravitational parameter is a constant as  
\begin{equation}\label{eq:gravitational_parameter}
\mu =G\left( m_{1}+m_{2}\right) ,
\end{equation}
where $G$, $m_{1}$, and $m_{2}$ are the gravitational constant, the mass of the primary (1.7~M$_{\odot}$), and the mass of the companion (0.2~M$_{\odot}$), respectively.

We use $V_{\text{proj}}$ to represent the projection of $V_{\bot}$ in the sky plane,
\begin{equation}\label{eq:v_proj}
V_{\text{proj} }=V_{\bot }\sqrt{\sin^{2} u+\cos^{2} u\cos^{2} i} .
\end{equation} 
The velocity in the sky plane that is perpendicular to the projected radial separation ($r_\text{proj}$) is 
\begin{equation}\label{eq:v_+_proj_tmp}
V_{\bot \text{proj} }=V_{\text{proj} }\frac{\cos i}{\sqrt{\sin^{2} u+\cos^{2} u\cos^{2} i} \sqrt{\cos^{2} u+\sin^{2} u\cos^{2} i} }.
\end{equation}
By substituting Equation~\eqref{eq:v_proj} into Equation~\eqref{eq:v_+_proj_tmp}, we obtain
\begin{equation}\label{eq:v_+_proj}
V_{\bot \text{proj} }=\frac{V_{\bot }\cos i}{\sqrt{\cos^{2} u+\sin^{2} u\cos^{2} i} } .
\end{equation}

The projected angular velocity of the companion in the sky plane is
\begin{equation}\label{eq:angular_velocity_sky_tmp}
\dot{u}_{\text{proj} } =\frac{V_{\bot \text{proj} }}{r_{\text{proj} }},
\end{equation}
which can be rewritten as 
\begin{equation}\label{eq:angular_velocity_sky}
\dot{u}_{\text{proj} } =\dot{u} \frac{\cos i}{\cos^{2} u+\sin^{2} u\cos^{2} i}
\end{equation}
by substituting Equation~\eqref{eq:r}, Equation~\eqref{eq:r_proj}, and Equation~\eqref{eq:v_+_proj} into Equation~\eqref{eq:angular_velocity_sky_tmp}. Equation~\eqref{eq:angular_velocity_sky} shows the projection of the angular velocity of the companion onto the sky plane.   

We adopted the orbits of the companion from \citet{Gonzalez2020}, which were derived from the astrometric fit to the companion positions listed in Table~\ref{table:astrometry}. For a given combination of the orbital elements, we first use Equation~\eqref{eq:r_proj} to numerically derive the true anomaly of the companion for a given orbit, adopting the separation $r_\text{proj}$ of 1.046\arcsec~in 2019. Based on Equation~\eqref{eq:r}, Equation~\eqref{eq:u}, Equation~\eqref{eq:angular_velocity}, and Equation~\eqref{eq:angular_velocity_sky}, we can derive the angular velocity of the companion in the sky plane using only the orbital elements (i.e., $a,e,\nu,\omega,i$). Finally, we deprojected the angular velocity of the companion in the sky plane to the disk plane, adopting the disk inclination, the disk position angle, and the companion position angle as $33\fdg81$, $144\fdg35$, and $133\fdg2$, respectively. The derived current angular velocity of the companion is $0\fdg457^{+0\fdg023}_{-0\fdg023}$~yr$^{-1}$. In Fig.~\ref{fig2}, we present the calculated maximum velocity of the companion in the disk plane, which is $0\fdg488^{+0\fdg355}_{-0\fdg154}$~yr$^{-1}$. The uncertainties are (16th, 84th) percentiles in Bayesian statistics.

We also validated the posterior probability distribution of orbital parameters by using $\texttt{orbitize!}$ \citep{orbitize_Blunt2017, orbitize_Blunt2020} and the same astrometric data shown in Table~\ref{table:astrometry}. Based on the new orbital parameters, we obtained the current angular velocity and maximum velocity of the companion to be $0\fdg419^{+0\fdg035}_{-0\fdg031}$~yr$^{-1}$ and $0\fdg560^{+0\fdg691}_{-0\fdg233}$~yr$^{-1}$, which is consistent with the results derived from \cite{Gonzalez2020}.

\section{Orbital parameters of the spiral-driving companion}
\begin{figure*}[h!]
        \includegraphics[width=0.99\textwidth]{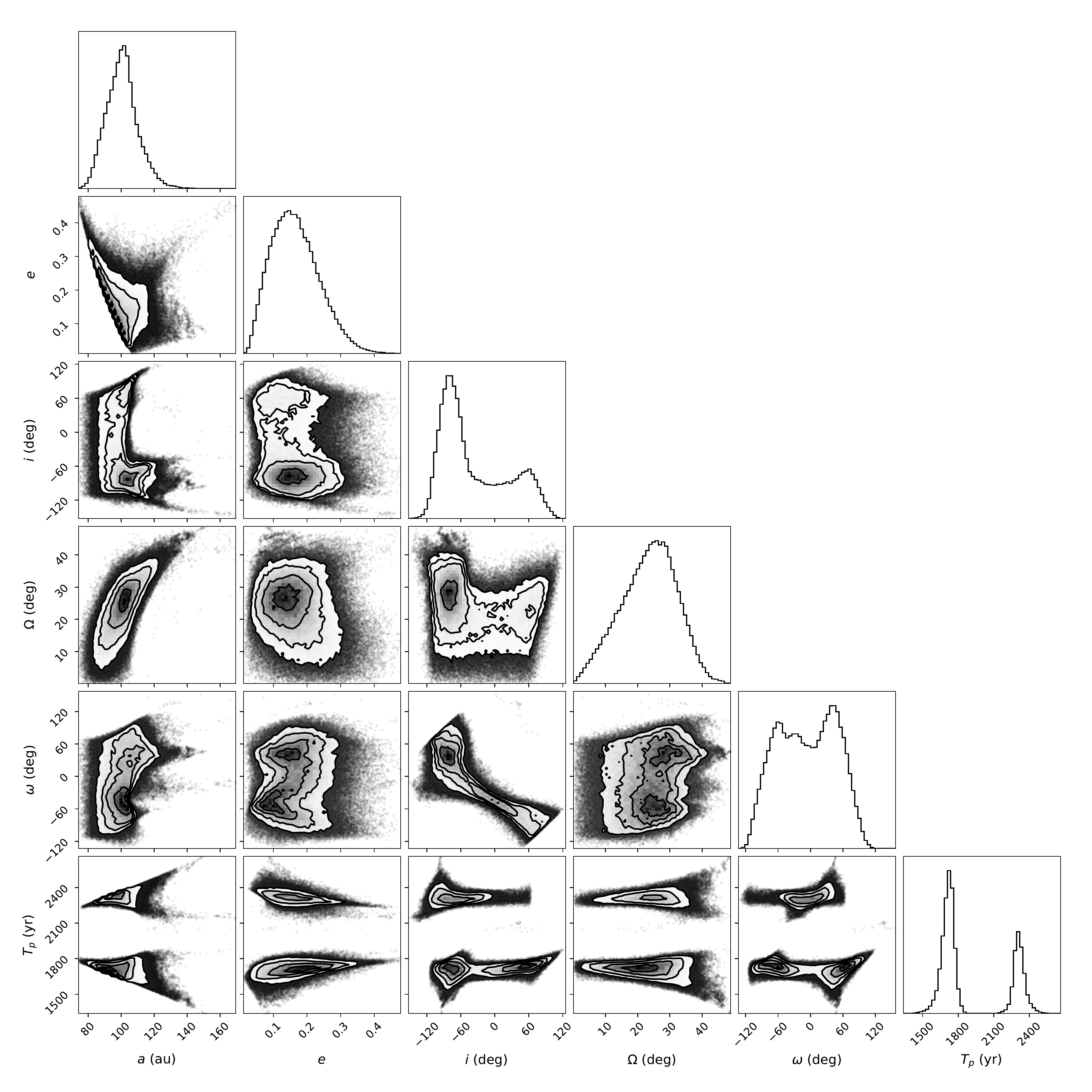}
    \caption{Distribution of orbital parameters that can dynamically drive the spiral.}
    \label{appendix_cornor_plot}
\end{figure*}

In general, the spiral pattern motion should be in the range of the slowest and fastest companion orbital frequency in the scenario of an eccentric perturber \citep[see Eq.~12 in][]{Zhu2022}.  From the posterior probability distribution of orbital parameters obtained by \cite{Gonzalez2020}, we derived the orbital parameters that satisfy Eq.~12 in \cite{Zhu2022}. In this case, the maximum orbital velocity of the companion is greater than the minimum spiral motion (0\fdg54 yr$^{-1}$). The corresponding distribution of orbital parameters is shown in Fig.~\ref{appendix_cornor_plot}.

\end{appendix}

\end{CJK*}
\end{document}